\documentstyle[12pt,aaspp4,flushrt]{article}
\newcommand{\etal} {{\it et~al.\ }}
\def\spose#1{\hbox to 0pt{#1\hss}}
\newcommand\lsim{\mathrel{\spose{\lower 3pt\hbox{$\mathchar"218$}}
     \raise 2.0pt\hbox{$\mathchar"13C$}}}
\newcommand\gsim{\mathrel{\spose{\lower 3pt\hbox{$\mathchar"218$}}
     \raise 2.0pt\hbox{$\mathchar"13E$}}}
\slugcomment{Submitted to {\it The Astrophysical Journal,} 22 March 1998}
\righthead{Radio Pulsar Masses}
\begin{document}
\title{Neutron Star Mass Measurements. I. Radio Pulsars}

\author{S. E. Thorsett\altaffilmark{1}}
\altaffiltext{1}{Alfred P. Sloan Research Fellow}
\affil{Joseph Henry Laboratories and Department of Physics, Princeton
  University, \\Princeton, NJ 08544; steve@pulsar.princeton.edu}

\and

\author{Deepto Chakrabarty\altaffilmark{2}}
\altaffiltext{2}{NASA Compton GRO Postdoctoral Fellow}
\affil{Center for Space Research, Massachusetts Institute of
  Technology, Cambridge, MA 02139;\\ deepto@space.mit.edu}
\begin{abstract}
  There are now about fifty known radio pulsars in binary systems,
  including at least five in double neutron star binaries. In some
  cases, the stellar masses can be directly determined from
  measurements of relativistic orbital effects. In others, only an
  indirect or statistical estimate of the masses is possible.  We
  review the general problem of mass measurement in radio pulsar
  binaries, and critically discuss all current estimates of the masses
  of radio pulsars and their companions. We find that significant
  constraints exist on the masses of twenty-one radio pulsars, and on
  five neutron star companions of radio pulsars.  All the measurements
  are consistent with a remarkably narrow underlying gaussian mass
  distribution, $m=1.35\pm0.04M_\odot$. There is no evidence that
  extensive mass accretion ($\Delta m\gsim0.1M_\odot$) has occurred in
  these systems. We also show that the observed inclinations of
  millisecond pulsar binaries are consistent with a random
  distribution, and thus find no evidence for either alignment or
  counteralignment of millisecond pulsar magnetic fields.
\end{abstract}
\section{Introduction}

Neutron stars have been the subject of considerable theoretical
investigation since long before they were discovered as astronomical
sources of radio and X-ray emission (\cite{bz34b,ov39,whe66}).  Their
properties are determined by the interplay of all four known
fundamental forces---electromagnetism, gravitation, and the strong and
weak nuclear forces---but neutron stars remain sufficiently simple in
their internal structure that realistic stellar modeling can be done.
Measurements of their masses and radii (as well as detailed study of
their cooling histories and rotational instabilities) provide a unique
window on the behavior of matter at densities well above that found in
atomic nuclei ($\rho_{\rm nuc}\approx
2.8\times10^{14}\mbox{g~cm}^{-3}$). Observations of neutron stars also
provide our only current probe of general relativity (GR) in the
``strong-field'' regime, where gravitational self-energy contributes
significantly to the stellar mass.

The most precisely measured physical parameter of any pulsar is its
spin frequency. The frequencies of the fastest observed pulsars
(PSR~B1937+21 at 641.9~Hz and B1957+20 at 622.1~Hz) have already been
used to set constraints on the nuclear equation of state at high
densities (e.g., \cite{fidp88}) under the assumption that these
pulsars are near their maximum (breakup) spin frequency.  However, the
fastest observed spin frequencies may be limited by complex accretion
physics rather than fundamental nuclear and gravitational physics. A
quantity more directly useful for comparison with physical theories is
the neutron star mass.  

The basis of most neutron star mass estimates is the analysis of
binary motion.  Soon after the discovery of the first binary radio
pulsar (\cite{ht75a}), it became clear that the measurement of
relativistic orbital effects allowed extremely precise mass estimates.
Indeed, the measurement uncertainties in several cases now exceed in
precision our knowledge of Newton's constant $G$, requiring masses to
be quoted in solar units $GM_\odot$ rather than kilograms if full
accuracy is to be retained.

After several recent pulsar surveys, there are now about fifty known
binary radio pulsar systems, of which five or six are thought to
contain two neutron stars.  It is thus possible for the first time to
consider compiling a statistically significant sample of neutron star
masses.  It is our purpose here to provide a general, critical review
of all current estimates of stellar masses in radio pulsar binaries.
The resulting catalog, with a careful, uniform approach to measurement
and systematic uncertainties, should be of value both to those who
wish to apply mass measurements to studies of nuclear physics, GR, and
stellar evolution, and as a guide to the critical observations for
observational pulsar astronomers. We begin with a discussion of known
methods for pulsar mass determination (\S\ref{sec:methods}), including
a new statistical technique for estimating the
masses of millisecond pulsars in non-relativistic systems.  In
\S\ref{sec:estimates} we review all known mass estimates, including new
data and analysis where possible.  Statistical analysis of the
available pulsar mass measurements is presented in \S\ref{sec:stat}.
We summarize in \S\ref{sec:summ}.

A second paper will consider mass estimates for neutron stars in X-ray
binary systems (\cite[Paper~II]{ct98}). A detailed discussion of the
implications of the combined results of this work and Paper~II for
studies of supernovae and neutron star formation, mass transfer in
binary evolution, the nuclear equation of state, and GR will occur
elsewhere (Paper~III).

\section{Methods of mass estimation}
\label{sec:methods}

It is a familiar circumstance that estimates of astronomical masses
are available only for bodies in gravitationally bound binary systems
or clusters. Compact stars introduce the additional possibility of
directly measuring the surface gravitational potential, and hence
$M/R$, through the study of redshifted spectral features. Although
this technique has been used with considerable success in the case of
white dwarfs, and attempts have been made to fit redshifted X-ray
spectra from neutron stars (Paper~II), no lines have been identified
in radio pulsar spectra and other gravitational effects on the
observed emission from pulsars are sufficiently complex and theory
dependent that no useful limits on the neutron star properties have yet
been possible. In the following, we thus limit ourselves only to
the determination of stellar masses in binary systems.

\subsection{Pulsar timing}

In any binary pulsar system, five Keplerian parameters can be very
precisely measured by pulse timing techniques (\cite{mt77}): the
binary period $P_b$, the projection of the pulsar's semimajor axis on
the line of sight $x\equiv a_1\sin i/c$ (where the binary inclination
$i$ is the angle between the line of sight and the orbital angular
momentum vector, defined to lie in the first quadrant), the
eccentricity $e$, and the time and longitude of periastron, $T_0$ and
$\omega_0$. It is frequently more convenient to use the orbital
angular frequency in place of the orbital period: $n\equiv2\pi/P_b$.
These observational parameters are related to the pulsar and companion
masses, $m_1$ and $m_2$, through the mass function
\begin{equation}
\label{eqn:massfunc}
f=\frac{(m_2\sin
  i)^3}{M^2}=n^2x^3\left(\frac{1}{T_\odot}\right)M_\odot,
\end{equation}
where $M\equiv m_1+m_2$, the masses are measured in solar units, and
we introduce the constant $T_\odot\equiv
GM_\odot/c^3=4.925\,490\,947\times10^{-6}$\,s.

Relativistic corrections to the binary equations of motion are most
often parameterized in terms of one or more post-Keplerian (PK)
parameters (\cite{dd86,tw89,dt92}).  In GR, the most significant PK
parameters have familiar interpretations as the advance of periastron
of the orbit $\dot\omega$, the combined effect of variations in the
transverse Doppler shift and gravitational redshift around an
elliptical orbit $\gamma$, the orbital decay due to emission of
quadrupole gravitational radiation $\dot P_b$, and the ``range'' and
``shape'' parameters $r$ and $s$ that characterize the Shapiro time
delay of the pulsar signal as it propagates through the gravitational
field of its companion.  In terms of measured quantities and the
pulsar and companion masses (in solar units), these PK parameters are
given by (\cite{tay92}):
\begin{eqnarray}
\label{eqn:PKomdot}
\dot\omega & = & 
3n^{\frac{5}{3}}\left(T_\odot
  M\right)^{\frac{2}{3}}\left(1-e^2\right)^{-1},\\
\gamma & = & en^{-\frac{1}{3}}T^{\frac{2}{3}}_\odot
M^{-\frac{4}{3}}m_2\left(m_1+2m_2\right),\\
\dot P_b & = &
-\frac{192\pi}{5}n^{\frac{5}{3}}
\left(1+\frac{73}{24}e^2+\frac{37}{96}e^4\right)
(1-e^2)^{-\frac{7}{2}}T_\odot^{\frac{5}{3}}m_1m_2M^{-\frac{1}{2}},\\
r & = & T_\odot m_2,\\ \label{eqn:PKs}
s & = & xn^{\frac{2}{3}}T_\odot^{-\frac{1}{3}}M^{\frac{2}{3}}m_2^{-1}.
\end{eqnarray}
Note, by combining equations~(\ref{eqn:massfunc}) and~(\ref{eqn:PKs}), that 
$s=\sin i$ for GR.)

The measurement of the mass function $f$ (equation~\ref{eqn:massfunc})
together with any two PK parameters
(equations~\ref{eqn:PKomdot}--\ref{eqn:PKs}) is sufficient in the context
of GR to uniquely determine the component masses $m_1$ and $m_2$.
With additional assumptions,
such as a uniform prior likelihood for orbital orientations with
respect to the observer (\S\ref{sec:uniform}), strong statements about
the posterior distribution of the masses is often possible if even a
single PK parameter is measured.

As an alternative to the PK formalism, which is designed for testing
gravitation theory, it is sometimes advantageous to fit the timing
data to a model that {\em assumes} the correctness of GR.  For
example, the DDGR model of Damour, Deruelle, and Taylor
(\cite{tay87a,tw89}) describes the pulsar phase as a function of the
five Keplerian parameters, the companion mass $m_2$, and the total
mass $M$. A least-squares fit of the timing data to the DDGR model
thus gives direct estimates of the uncertainties in $m_2$ and $M$, as
well as the covariances between the mass estimates and estimates of
other parameters.

It is of interest to note that mass measurements from pulsar timing
observations depend on the unknown relative motion of the solar system
and the binary barycenters.  Damour and Deruelle (1986)\nocite{dd86}
have shown that neglecting this velocity is equivalent to changing
units of mass and time.  In particular, the rest frame mass $m$ and
the barycenter frame mass $m^{\rm ssb}$ are related by $m=Dm^{\rm
ssb}$, with Doppler factor
\begin{equation}\label{eqn:Doppler}
D\equiv\frac{1-{\mathbf{\hat n}}\cdot{\mathbf{v}}_b/c}{\sqrt{1-v^2_b/c^2}},
\end{equation}
where $\mathbf{\hat n}$ is the line of sight unit vector and
${\mathbf{v}}_b$ is the barycentric velocity of the pulsar. 
Although the transverse velocity of the binary system can be estimated
from proper motion measurements, the radial component
${\mathbf{\hat n}}\cdot{\mathbf{v}}_b$ is unknown. For a typical velocity of
100~km~s$^{-1}$, the systematic mass error is about $\sim0.03\%$;
small, but in some cases much larger than other uncertainties.

\subsection{Masses of companion stars}

While timing measurements of relativistic corrections to the Keplerian
orbital equations provide the most accurate and theory-independent
estimates of neutron star masses, they are possible only for close,
eccentric binary orbits or when the orbit is observed nearly edge-on.
In the great majority of observed binaries, the mass function provides
the only timing information about the component masses, and the pulsar
mass can only be determined if additional constraints are found on
the companion mass and binary inclination through other techniques.
This section describes several ways to use observations or theoretical
considerations to limit $m_2$; alternate limits on $\sin i$ are the
topic of \S\ref{sec:sini}.

\subsubsection{Optical observations of white dwarf companions}

In recent years, over a dozen companion stars in radio pulsar binaries
have been optically detected (e.g., \cite{van96,lfc96}). In most
cases, the companions are white dwarfs; the two main sequence
exceptions are discussed in the next section.  Many of the white dwarf
companions are extremely faint ($m_V\sim26$), allowing detection only
with the {\em Hubble Space Telescope } and the largest ground-based
telescopes.

There are a number of ways to determine white dwarf masses (e.g.,
\cite{reid96} and references therein). Given a theoretical relation
between the white dwarf mass and radius, the measurement of any
combination of the mass and radius is sufficient to determine the
mass. For example, the radius can be estimated directly from estimates
of the optical flux, effective temperature, and distance.
Alternately, the surface gravity, $\log g$, can be found by fitting a
model atmosphere to the observed spectrum (\cite{bwf91}), and
combining the result with a temperature or luminosity estimate.  In
practice, difficulties arise from several sources.  First, the white
dwarf companions of millisecond pulsars, with typical masses
$m_2<0.5M_\odot$, are usually believed to be helium white dwarfs which
were insufficiently massive to burn to carbon; the luminosity and
temperature evolution of such stars has received much less study than
for more massive white dwarfs (e.g., \cite{dm90}), leading to
uncertainties in the finite temperature contributions to the
mass-radius relationship. Hansen and Phinney
(1998a,b)\nocite{hp98a,hp98b} have recently calculated cooling curves
for helium core dwarfs, using their own calculations of H and He
opacities at temperatures below 6000~K, and applied their results to a
number of radio pulsar companions. We discuss their results in more
detail in \S\ref{sec:estimates}.

The measurement of surface gravity in cool stars is also potentially
problematic. A higher helium abundance, as may occur through enhanced
convective mixing in cool stars, will produce higher pressure and
broader lines and hence mimic a higher white dwarf mass (\cite{bwf91});
indeed, there is some evidence that surface gravity measurements
overestimate masses below about 12,000~K (\cite{reid96}).

\subsubsection{Optical observations of main sequence companions}

Two pulsars have been found in binaries with main sequence companions:
PSR~B1259$-$63, with a Be companion (\cite{jml+92}), and
PSR~J0045$-$7319, with a B star companion (\cite{kjb+94}).  Although
in both cases the optical companion is quite bright and easily
observed, knowledge of the companion mass $m_2$ and pulsar mass
function $f$ is of little use in limiting $m_1$ when $m_1\ll m_2$. If
the mass function of the companion can also be measured, then the mass
ratio can be determined.  This has been done in the case of
J0045$-$7319 (\S\ref{sec:0045}).

\subsubsection{The $P_b$--$m_2$ relation}
\label{sec:cmrr}

The binary millisecond pulsars are believed to be spun up to high spin
frequencies through mass transfer from a companion star. In many
cases, these pulsars are observed in wide, low-eccentricity binary
systems with white dwarf secondaries.  These characteristics indicate
that the secondary must have passed through a red giant phase after
the formation of the neutron star primary, during which tidal torques
circularized the orbit and the giant probably filled its Roche lobe,
causing the mass transfer which spun up the pulsar to millisecond
periods.  At the end of mass transfer, the envelope of the giant is
exhausted or ejected, leaving the degenerate core as a white dwarf
secondary.  There is a close relation between the core mass and the
radius of low mass giants (\cite{rw71,wrs83,jrl87,rpj+95,rj97}).
Combined with the assumption that the giant filled its Roche lobe during
the mass transfer, this yields a relation between the binary period at
the end of mass transfer and the remnant white dwarf mass
(\cite{rpj+95}).  

Following Rappaport \etal (1995) and references therein, the relation
between the effective radius of the Roche lobe $R_L$ and the binary
separation $a$ can be written
\begin{equation}
R_L\approx 0.46\, a\, \left(1+\frac{m_1}{m_g}\right)^{-1/3},
\end{equation}
where $m_g$ is the total mass of the giant (core and envelope).
Near the end of mass transfer, the envelope mass can be neglected, so
that $m_g\approx m_2$ (where $m_2$ is the final white dwarf mass), and
the giant radius $R_g\approx R_L$. Using Kepler's third law,
\begin{equation}
\label{eqn:pb}
P_b=0.374\, R_g^{3/2}\, m_2^{-1/2}\quad\mbox{days},
\end{equation}
where $R_g$ and $m_2$ are in solar units. We note that
equation~(\ref{eqn:pb}) is independent of the pulsar mass $m_1$, and relates
the orbital period at the end of mass transfer to the final white
dwarf mass and the giant radius $R_g$. The radius depends, in general,
on the composition and history of the giant as well as on $m_2$; the
utility of equation~(\ref{eqn:pb}) derives from the relatively narrow
distribution of $R_g$ for a given $m_2$.

Rappaport \etal (1995)\nocite{rpj+95} have examined stellar models
with a wide range of chemical compositions, varying between Population~I and
Population~II values, as well as a variety of envelope masses and convective
mixing lengths. Over a wide range of core masses greater than
$0.15M_\odot$, they find the data are well described by the equation
\begin{equation}
\label{eqn:rapp1}
R_g=\frac{R_0\, m_2^{4.5}}{1+4\, m_2^4}+0.5
\end{equation}
where $R_0=4950R_\odot$ is the best fit to the stellar models.  In all
models studied by Rappaport \etal, equation~(\ref{eqn:rapp1}) was
correct within a factor of 1.8.  Over the limited range
$m_2<0.25M_\odot$, Rappaport and Joss (1997)\nocite{rj97} found that
the alternate expression
\begin{equation}
\label{eqn:rapp2}
\log R_g=0.031+1.718\, m_2+8.04\, m_2^2
\end{equation}
could be used, with the smaller uncertainty of a factor $\sim 1.3$. We 
regard these to be approximately 95\% confidence regions for $R_g$.

It is important to determine the range of applicability of
equations~(\ref{eqn:rapp1}) and~(\ref{eqn:rapp2}).  At orbital periods
less than a few days, heating of the surface of the companion by
X-rays produced during accretion may cause significant bloating and
modification of the core mass--radius relations (\cite{pod91,rpj+95}).
Following Rappaport {\it et al.,} we trust the modeling only for
binary orbits longer than three days. Further, the $P_b$--$m_2$
relation cannot be applied to systems like PSR~J2145$-$0750 or
J1022+10---both 16~ms pulsars that were most likely recycled in common
envelopes that occurred when their companions overflowed their Roche
surfaces while on the asymptotic giant branch, leaving carbon-oxygen
rather than helium white dwarfs (\cite{vdh94})---or to pulsars like
B0820+02, a slow (0.864~s), high-field ($3\times10^{11}$~G) pulsar
without evidence of significant recycling.  To avoid both classes of
pulsar without introducing biases by cutting on apparent companion
mass, we apply the $P_b$--$m_2$ relation only to pulsars with
$P<10$~ms, for which the assumption of an extended period of mass
transfer during a low-mass X-ray binary phase seems secure.

Using equations~(\ref{eqn:pb})--(\ref{eqn:rapp2}), the observed orbital period
in a millisecond pulsar binary can be used to limit the range of
$m_2$. When combined with the mass function (equation~\ref{eqn:massfunc})
and the restriction $\sin i<1$, an upper limit can be placed on the
pulsar mass. The resulting limits are discussed below, in
\S\ref{sec:nswd}. If a random distribution of orbits is assumed
for the set of observed millisecond pulsars, then statistical
arguments lead to limits on the masses of milliseconds pulsars as a
class. These arguments are described in \S\ref{sec:stat}.

\subsection{Alternative methods of estimating the orbital inclination}
\label{sec:sini}

If the companion mass can be estimated and bounds can be found on the
orbital inclination, then the mass of the pulsar can be found using
the Keplerian mass function.  In this section, we discuss 
measurements that may allow direct estimates of $\sin i$.

\subsubsection{Polarization measurements}
\label{sec:polar}

In the standard model of millisecond pulsar formation, the pulsar is
spun-up by mass transfer from a companion star.  After spin-up, the
pulsar spin axis will be aligned with the orbital angular momentum, so
a measurement of the angle $\zeta$ between the pulsar spin axis and
the line of sight also determines the orbital inclination: $i=\zeta$,
or $i=\pi-\zeta$ for $\zeta$ in the second quadrant.
(Although this is also true at an early stage of the formation of
double neutron star binaries, an asymmetry in the second supernova
explosion may leave the spin and orbit misaligned.)

In the rotating vector model of pulsar emission, the radio signal is
elliptically polarized with the major axis of the polarization ellipse
aligned with the plane of curvature of the dipolar magnetic field
lines. The position angle of the observed linear polarization $\psi$
can be expressed as a function of the rotational phase $\phi$ of the
pulsar:
\begin{equation}
\label{eqn:rvm}
\tan\left(\psi(\phi)-\psi_0\right)=
\frac{\sin\alpha\sin\phi}{\sin\zeta\cos\alpha-\cos\zeta\sin\alpha\cos\phi},
\end{equation}
where $\alpha$ is the angle between the pulsar spin axis and the
magnetic pole.  In practice, the difference $\zeta-\alpha$ can often
be estimated quite well from equation~(\ref{eqn:rvm}), but accurate
estimates of $\zeta$ or $\alpha$ are possible only if polarized
emission can be detected over a broad range of pulse phase $\phi$.

Application of equation~(\ref{eqn:rvm}) to millisecond pulsars, in which the
light cylinder bounds the magnetosphere at only a few stellar radii,
may be complicated by multipolar field geometries, aberration, and
other deviations from the simple rotating vector model. Indeed,
attempts to fit equation~(\ref{eqn:rvm}) to millisecond pulsar data
have met with few unqualified successes (\cite{ts90,nms+97}).

\subsubsection{Interstellar Scintillation}
\label{sec:scint}

Observed pulsar signal strengths vary in both frequency and time
because of scintillation in the interstellar medium.  A phase-changing
screen along the line of sight produces an interference pattern across
which the pulsar moves; the characteristic scintillation decorrelation
timescale $\tau_{\rm iss}$ is inversely proportional to the transverse
component of the pulsar velocity. Observations of scintillation rates
have been widely used to estimate the proper motions of isolated
pulsars (e.g., \cite{cor86}).  In binary systems, the transverse
velocity is modulated by the orbital motion, with a small amplitude of
modulation for orbits viewed face-on, and a large amplitude for orbits
viewed on edge.  Scintillation measurements can thus provide an
alternate path to estimating the inclination angle $\sin i$
(\cite{lyn84}).

The real physical situation is more complex.  The measured
scintillation velocity ${\mathbf{v}}_{\rm iss}$ of a pulsar depends not
only on its proper motion ${\mathbf{v}}_{\rm pm}$ and orbital motion
${\mathbf{v}}_{\rm orb}(t)$, but also on the Earth's motion
${\mathbf{v}}_{\earth}(t)$, the mean motion of the scattering medium
${\mathbf{v}}_{\rm ism}$ and the ratio of the effective distance to the
scattering screen to the pulsar distance, $f$:
\begin{equation}
{\mathbf{v}}_{\rm iss}(t)=\left(1-f\right)\left({\mathbf{v}}_{\rm
  orb}(t)+{\mathbf{v}}_{\rm
  pm}\right)+f{\mathbf{v}}_{\earth}(t)-{\mathbf{v}}_{\rm ism}.
\end{equation}
If the proper motion is known, from timing measurements or
interferometry, then the annual modulation of ${\mathbf{v}}_{\earth}$
and the orbital modulation of ${\mathbf{v}}_{\rm orb}$ can in principle
be used to find not only the orbital inclination $\sin i$ but also
$f$, the pulsar distance $d$, and the position angle on the sky of the
orbital ascending node $\Omega$.

Only a few attempts to determine orbital inclinations from
scintillation observations have been published.  In no case was
${\mathbf{v}}_{\rm pm}$ known, nor has the annual modulation due to
${\mathbf{v}}_{\earth}$ been measured (although the latter has been
observed in the isolated millisecond pulsar B1937+21 by Ryba
(1991)).\nocite{ryb91} Lyne (1984)\nocite{lyn84} found that the
inclination of PSR~B0655+64 is either $62^\circ$ or $84^\circ$ (the
discrete ambiguity could be broken through observations at another
time of year, using variations in ${\mathbf{v}}_{\earth}$). Jones and
Lyne (1988)\nocite{jl88} later expressed reservations about this
measurement.  In the most convincing success of the scintillation
technique, Dewey \etal (1988)\nocite{dcww88} found a limit on the the
inclination of PSR~B1855+09: $\sin i\ge 0.94$. Shapiro time delay
measurements have since shown that $\sin i\approx0.9993$
(\S\ref{sec:1855}).

The problem of estimating scintillation parameters from data is
considered by Cordes (1986).\nocite{cor86} The dominant error source
in estimating scintillation parameters is the finite number $N$ of
scintillation features sampled: $\sigma_{v}/v_{\rm
  iss}\approx0.6/\sqrt{N}$.  Scintillation intensity fluctuations are
exponentially distributed; for observing bandwidth $B$ and time $T$,
we have roughly $N\approx10^{-2}BT/(\Delta\nu_{\rm iss}\tau_{\rm
  iss})$. Considering a case relevant to many recently discovered
pulsar binaries, a pulsar at a dispersion measure DM$=20$ will have
typical scintillation parameters at 430~MHz $\tau_{\rm
  iss}\sim10$~minutes and $\Delta\nu_{\rm iss}\sim100$~kHz
(\cite{ric88}).  With $B=10$~MHz and $T=1$~hr, $\sigma_v/v_{\rm
  iss}\sim25\%$, so a study of orbital dependence of scintillation
parameters requires a substantial amount of observing time.
Furthermore, to use a bandwidth of 10~MHz requires a spectrometer with
$\gsim100$ frequency channels. Such observations will be more easily
done with the new generation of flexible, all-digital pulsar data
recorders (\cite{ssd+96,jcpu97}). It is possible that long
term variability in $f$ or ${\mathbf{v}}_{\rm iss}$ will limit the use
of the annual variation in ${\mathbf{v}}_{\earth}$ to estimate $f$.

\subsubsection{Secular variation of $x=a_1\sin i$}
\label{sec:xdot}

Proper motion of the binary system across the sky leads to a secular
change in the projected semimajor axis $x$. If $\Omega$ is the
position angle of the ascending node and $\mu_\alpha$ and $\mu_\delta$
are the components of the proper motion \mbox{\boldmath$\mu$} in right
ascension and declination, then
\begin{equation}
\label{eqn:xdot}
\frac{\dot x}{x}=\cot i\left(-\mu_\alpha\sin\Omega+\mu_\delta\cos\Omega\right).
\end{equation}
The angle $\Omega$ is generally unknown (though it is accessible to
scintillation measurements, \S\ref{sec:scint}), but
eqn.~\ref{eqn:xdot} can be rewritten as the limit
\begin{equation}
\tan i<\left|\frac{x}{\dot x}\mu\right|.
\end{equation}
A similar expression can be found relating proper motion to a
purely geometric advance of periastron $\dot\omega$---in principle,
measurement of both $\dot x$ and $\dot\omega$ would determine $i$---but the
effect is too small to have yet been seen in any pulsar binary.

\subsubsection{Random distribution of orbital inclinations}
\label{sec:uniform}

When no other information is available on the orbital inclination, it
is sometimes useful in statistical calculations to assume that binary
orbits are randomly oriented with respect to the line of sight. The
differential distribution of inclinations is then proportional to
$\sin i$ (i.e., the most likely orbital viewing angle is
edge-on). Values of $\cos i$ should then be uniformly distributed
between 0 and 1.


In binary systems where the directions of the pulsar spin axis and the
orbital angular momentum vector are expected to be correlated, such as
millisecond pulsar systems in which significant mass transfer has
occurred since the last supernova, non-random orientations of the
pulsar magnetic field axis with repect to the spin axis may lead to
a pulsar discovery bias that skews the observed orbital orientation
distribution. Some models of pulsar evolution predict magnetic field
alignment, counteralignment, or both (e.g., \cite{rud91}). 

Backer (1998) has claimed, based on the distribution of observed mass
functions in 21 pulsar--white-dwarf binaries, that millisecond pulsars
have a non-random distribution of observed binary inclinations. He
suggests that an apparent preference for high inclination orbits may
arise because the pulsars' magnetic fields are preferentially oriented
perpendicular to the spin axis.  However, his analysis makes several
assumptions and approximations that prove unwarranted.  In particular,
he compares the observed distribution of minimum companion masses
$m_{\rm 2,min}$ with the predicted distribution given the observed
mass functions, random inclinations and a single fixed value for the
true companion masses $m_2$ (i.e., a delta function distribution in
$m_2$), finding poor agreement.  However, the predicted distribution
for $m_{\rm 2,min}$ actually depends quite sensitively upon the
assumed distribution for $m_2$; the delta function distribution is an
unreasonable assumption.  Indeed, the $P_b$--$m_2$ relation suggests
that the observed $m_2$ values should vary by a factor of three.
Knowing the precise limits and distribution of these masses is crucial
for analyzing the distribution of $m_{\rm 2,min}$.

We repeated Backer's analysis, with different assumptions about the
distribution of $m_2$ values.  If we assume the true values $m_2$ fall
uniformly in the range 0.15--0.35~$M_\odot$, then the predicted
and observed distributions for $m_{\rm 2,min}$ agree at an 81\%
confidence level according to a Kolmogorov-Smirnov (KS) test (e.g.,
\cite{edj+71}).  However, assuming a 0.1--0.4~$M_\odot$ range reduces
the agreement to the 33\% confidence level.  The discrepancy is
primarily due to a deficit of observed systems with small values of
$m_{\rm 2,min}$, as noted by Backer.  Unfortunately, our knowledge of the
true underlying distribution of $m_2$ in these binaries is
sufficiently uncertain that no conclusion regarding the distribution
of inclinations is possible from this line of analysis.
However, we argue below (\S\ref{sec:stat}) that for systems with
orbital periods longer than three days, where the $P_b$--$m_2$
relation can be applied to estimate the $m_2$ distribution, the
observed mass functions are consistent with a random distribution of
inclinations. 

\section{Radio pulsar masses}
\label{sec:estimates}

The parameters of 47 radio pulsar binary systems are listed in
Table~\ref{tab:allbins}.\placetable{tab:allbins} At least five of
these pulsars have neutron star companions, so a total of 52 neutron
stars are known in radio pulsar binaries. Useful mass constraints can
be placed on half of these stars.

\subsection{Double neutron star binaries}
\label{sec:nsns}

\subsubsection{PSR~J1518+4904}

PSR~J1518+4904 is in a moderately relativistic binary system; one PK
parameter has been measured, the relativistic advance of periastron
$\dot\omega=0.0111(2)^\circ\mbox{yr}^{-1}$, yielding a total system
mass of $2.62(7)M_\odot$ (\cite{nst96}).  Given the mass function
$f=0.115988M_\odot$, the lower limit on the companion mass (using
$\sin i<1$) is $m_2>(fM^2)^{1/3}=0.93M_\odot$, and the lower limit on
the inclination (given $m_1>0$) is $\sin i>(f/M)^{1/3}=0.35$ (or
$i>20^\circ$).  Using a uniform prior distribution in $\sin i$ over
the interval $0.35<\sin i<1$, we find the central 68\% confidence
intervals $m_1=1.56^{+0.13}_{-0.44}$ and $m_2=1.05^{+0.45}_{-0.11}$
and central 95\% confidence intervals $m_1=1.56^{+0.20}_{-1.20}$ and
$m_2=1.05^{+1.21}_{-0.14}$.

The identification of the companion as a neutron star is compelling,
although it is not possible to completely rule out a low-mass black
hole. Optical observations (van~Kerkwijk, quoted in Sayer
1996)\nocite{say96} show no source at the pulsar position to
$m_B\sim24.5$, excluding a main sequence companion, and a white dwarf
in an eccentric orbit is not expected on evolutionary grounds.

\subsubsection{PSR~B1534+12}

PSR~B1534+12 is in a highly relativistic binary system, presumably
with a second neutron star companion. All five of
the PK parameters defined by equations~(\ref{eqn:PKomdot})--(\ref{eqn:PKs})
have been measured, three with better than 1\% precision
(\cite{sac+98}).  Using the DDGR timing model, the total system mass
is found to be $M=2.67838(8)M_\odot$, while the individual
component masses $m_1$ and $m_2$ are both $1.339(3)M_\odot$.
(The quoted errors are 68\% confidence regions; the 95\% confidence
regions are about twice as large.) It is remarkable that the pulsar
and companion masses agree to better than $0.4\%$; the assumption that
the companion is a second neutron star seems secure. The uncertainty
on the individual masses is expected to decrease significantly in the
next year, when observations are again possible with the Arecibo
telescope.

\subsubsection{PSR~B1913+16}

PSR~B1913+16 was the first binary pulsar discovered and, with
observations stretching over two decades, it remains one of the best
studied. PSR~B1913 is a highly relativistic system: the PK parameters
$\dot\omega$, $\gamma$, and $\dot P_b$ have all been measured
precisely.  A fit to the DDGR timing model yields the total system
mass $M=2.82843(2)$ and component masses $m_1=1.4411(7)$ and
$m_2=1.3874(7)M_\odot$ (\cite{tay92}). (The quoted errors are 68\%
confidence regions; the 95\% confidence regions are about twice as
large.)  The uncertainties are approaching the level where they will
be dominated by kinematic effects (equation~\ref{eqn:Doppler}).

\subsubsection{PSR~B2127+11C}

PSR~B2127+11C is a relativistic binary in the globular cluster M15.
Precise measurements have been made of the PK parameters $\dot\omega$
and $\gamma$, resulting in mass estimates for the pulsar and companion
of $1.349(40)$ and $1.363(40)M_\odot$, respectively, and a total
system mass $M=2.7121(6)M_\odot$ (\cite{dk96}).  (The quoted errors
are 68\% confidence regions; the 95\% confidence regions are
approximately twice as large.)

\subsubsection{PSR~B2303+46}

PSR~B2303+46 is a pulsar in a moderately relativistic orbit with, most
likely, a neutron star companion. One PK parameter, the relativistic
advance of periastron, has been measured.  A timing analysis was
published by Thorsett \etal (1993)\nocite{tamt93}, and updated by
Arzoumanian (1995)\nocite{arz95}. Using data from early 1985 to late
1994, we find an improved value
$\dot\omega=0.01019(13)^\circ\mbox{yr}^{-1}$, yielding a total system
mass $M=2.64\pm0.05M_\odot$ and the constraints $m_1<1.44M_\odot$ and
$m_2>1.20M_\odot$.  Using a uniform prior distribution in $\sin i$, we
find the central 68\% confidence intervals $m_1=1.30^{+0.13}_{-0.46}$
and $m_2=1.34^{+0.47}_{-0.13}$ and central 95\% confidence intervals
$m_1=1.30^{+0.18}_{-1.08}$ and $m_2=1.34^{+1.08}_{-0.15}$.

\subsection{Neutron star/white dwarf binaries}
\label{sec:nswd}

\subsubsection{PSR J0437$-$4715}

PSR~J0437$-$4715 is the brightest and closest known millisecond
pulsar, and therefore one of the best studied. It has a mass function
$f=1.243\times10^{-3}M_\odot$, and a white dwarf companion that has
been detected optically (\cite{jlh+93,bai93,bbb93,dbd93}); and both
thermal and nonthermal X-ray emission from the neutron star have been
observed (\cite{bt93a,hmm96}).  The distance is known,
$d=178\pm26$~pc, from the effects of parallax on the pulsar timing
signal (\cite{sbm+97}).  Using the optical data of Danziger {\it et al.,}
Hansen and Phinney (1998b)\nocite{hp98b} found an effective
temperature of the companion $T_{\rm eff}=4600\pm200$~K.  Combining
the $T_{\rm eff}$ and distance estimates with their cooling curves,
they find consistent companion models for all masses $0.15<m_2<0.375
M_\odot$.  This range encompasses the mass limits derived from the
$P_b$--$m_2$ relation (\S\ref{sec:cmrr}), $0.16<m_2<0.23 M_\odot$.

The binary orbit is nearly circular, and despite the very high timing
precision achieved, no PK timing parameters have been measured.
However, the proximity of the pulsar leads to a high proper motion
($\mu=141$~mas/yr), changing the projected orbital size
(\S\ref{sec:xdot}) at a rate $\dot x/x=2.43(12)\times 10^{-14}$
(\cite{sbm+97}).  The implied limit on the inclination angle is
$i<43^\circ$, or $\sin i<0.682$.

Attempts to measure the system geometry using the polarization of the
radio beam (\S\ref{sec:polar}) have proven difficult, because of the
very complex emission pattern. Well-calibrated intensity and
polarization data have been reported by Manchester and Johnston
(1995)\nocite{mj95}, who have modeled the sweep of the linear
polarization position angle across the profile in terms of the
rotating vector model. They find that an impact parameter of the line
of sight on the magnetic pole $\beta=-5^\circ$ and an angle between
the line of sight and the spin axis $\zeta=140^\circ$ produced
reasonable agreement with the data, but there are strong systematic
deviations from the simple rotating vector model. Using the same data,
Gil and Krawczyk (1997)\nocite{gk97} model the multicomponent profile
and find a comparable impact parameter, $\beta=-4^\circ$, but a very
different $\zeta=16^\circ$. They show that this geometry also explains
the observed polarization sweep over much of the pulse period.  With
the assumption that the pulsar spin axis is aligned with the orbital
axis, the implied orbital inclination is $\sin i=0.64$ (Manchester and
Johnston model) or $\sin i=0.28$ (Gil and Krawczyk model).  Either is
consistent with the timing data.

The timing limit on $\sin i$, together with the mass function and the
upper limit on $m_2$ from the core mass--orbital period relation,
gives an upper limit to the pulsar mass of $m_1<1.51M_\odot$, while
the optical upper limit on $m_2$ gives only the much weaker limit
$m_1<3.29M_\odot$.  The Manchester and Johnston inclination together
with the $P_b$--$m_2$ mass range gives $0.77<m_1<1.37M_\odot$, while
the Gil and Krawczyk inclination gives $0.11<m_1<0.23M_\odot$, a
physically implausible result.  In any case, the complexity of the
pulse shape modeling leads us to adopt the weaker one-sided limit
$m_1<1.51M_\odot$.

Improvements of the parallax measurement and the optical photometry
and spectroscopy are needed to further constrain the companion mass.
An independent mass determination would test the core mass--orbital
period relation as well as improve the limit on $m_1$.  Measurement of
the position angle of the ascending node (perhaps through
scintillation studies), combined with the measurement of $\dot x/x$,
would give the inclination of the orbit.

\subsubsection{PSR~J1012+5307}

PSR~J1012+5307 has a hot, bright white dwarf companion that has been
extensively studied.  van Kerkwijk, Bergeron, and Kulkarni
(1996)\nocite{vbk96} have used the models of Bergeron, Wesemael, and
Fontaine (1991)\nocite{bwf91} to determine the effective temperature
$T_{\rm eff}=8550\pm25$~K and surface gravity $\log g=6.75\pm0.07$ of
the companion. The latter value is in disagreement with an unpublished
value of Callanan and Koester, $\log g=6.4\pm0.2$, quoted by Hansen
and Phinney (1998b)\nocite{hp98b}.
Using the optical observations and their cooling models for low mass
helium white dwarfs, Hansen and Phinney find a companion mass
$0.165<m_2<0.215M_\odot$ for the van~Kerkwijk \etal gravity
measurement and $0.13<m_2<0.18$ for the Callanan and Koester
value.

The radial velocity of the companion has been measured, making
J1012+5307 a double-line spectroscopic pulsar binary.
van~Kerkwijk \etal have found $m_1/m_2=9.5\pm0.5$ (van~Kerkwijk,
private communication; their earlier published value
$m_1/m_2=13.3\pm0.7$ was corrupted by a calibration problem).
Depending on the gravity measurement, the resulting ($1\sigma$) pulsar
mass limit is $1.5<m_1<2.2M_\odot$ or $1.2<m_1<1.8M_\odot$. Clearly
resolution of the discrepant gravity measurements must be a high
priority, the radial velocity measurements now contribute negligibly
to the total error on $m_2$.  For now, we adopt Hansen and Phinney's
conservative conclusion that $1.2<m_1<2.2M_\odot$, and we regard the
error range as roughly a 68\% confidence interval.

\subsubsection{PSR~J1045$-$4509}

PSR~J1045$-$4509 is in a 4.08~day orbit with mass function
$1.765\times10^{-3}M_\odot$. The $P_b$--$m_2$ relation gives an upper
limit to the companion mass of $0.168M_\odot$, leading to a limit on
the pulsar mass of $m_1<1.48M_\odot$.

\subsubsection{PSR~J1713+0747}

PSR~J1713+0747 is a bright pulsar in a nearly circular, 68~day orbit
with a white dwarf companion. At the timing precision reached (about
500~ns for 1.4~GHz observations), Camilo (1995)\nocite{cam95a} found
that the Shapiro PK parameters $r$ and $s$ were required to adequately
model the data, but strong covariances between the range parameter $r$
and other orbital parameters (especially $x$) prevented him from
setting interesting limits on the pulsar or companion mass.

The $P_b$--$m_2$ relation predicts $0.26 <m_2< 0.35 M_\odot$.
Companion mass estimates are also possible by combining 
unpublished optical observations of the white dwarf by Lundgren \etal
with the parallax from pulsar timing (\cite{cfw94}).  The resulting
limits are $0.15<m_2<0.31M_\odot$ if the white dwarf has a thick H
envelope, and $m_2<0.27M_\odot$ if it has a thin H envelope
(\cite{hp98b}).

If the observed mass function $f=7.896\times10^{-3}M_\odot$ is
combined with the $P_b$--$m_2$ limit on $m_2$ and the restriction
$\sin i<1$, an upper limit to the pulsar mass is found:
$m_1<1.94M_\odot$. The optical observations yield a tighter limit,
$m_1<1.63M_\odot$, as was noted by Hansen and Phinney.  More
interesting limits can be obtained by combining the Shapiro
measurements with the limits on $m_2$ (recalling that $m_2=r$). For
example, if $m_2=0.299M_\odot$, the timing data of Camilo is
consistent with only the restricted range $\sin i=0.963(4)$.  By
letting $m_2$ vary over the range allowed by the $P_b$--$m_2$
relation, we find the allowed pulsar mass $m_1=1.45\pm0.31$, where the
uncertainties are dominated by the systematic uncertainties in the
$P_b$--$m_2$ relation. If the companion mass is limited to the
intersection of the regions allowed by $P_b$--$m_2$ and the optical
measurements, $0.26<m_2<0.31M_\odot$, then we find
$m_1=1.34\pm0.20M_\odot$.

\subsubsection{PSR~B1802$-$07}

PSR~B1802$-$07 is in the globular cluster NGC~6539. Its companion is
most likely a white dwarf; the system's large eccentricity can be
understood as the result of gravitational perturbations of the system
by close stellar encounters in the dense cluster.

The relativistic advance of periastron $\dot\omega$ was measured by
Thorsett \etal (1993)\nocite{tamt93}, and an improved measurement was
published by Arzoumanian (1995)\nocite{arz95}. Using the same analysis
techniques, with data extending through October 1997, we find a
slightly improved value $\dot\omega=0.0578(16)$, implying a total
system mass $M=1.62(7)M_\odot$. The lower bound on the companion mass
from the requirement that $\sin i<1$ is $(fM^2)^{1/3}=0.29M_\odot$,
and the lower bound on the inclination from the requirement $m_1>0$ is
$\sin i>(f/M)^{1/3}=0.18$ (or $i>10^\circ$).  Using a uniform prior
for $\sin i$ in the range $0.18<\sin i<1$, we find a 68\% confidence
bound on the pulsar mass $m_1=1.26^{+0.08}_{-0.17}M_\odot$ and a 95\%
confidence bound $m_1=1.26^{+0.15}_{-0.67}M_\odot$.

\subsubsection{PSR~J1804$-$2718}

PSR~J1804$-$2718 is in an 11.1~day orbit.  The $P_b$--$m_2$ relation
limits its companion mass to $0.185 <m_2< 0.253M_\odot$; the resulting
upper limit on the pulsar mass is $m_1<1.73M_\odot$.

\subsubsection{PSR~B1855+09}
\label{sec:1855}

PSR~B1855+09 is a 5.4~ms pulsar in a 12.3~day circular orbit
($e=2\times10^{-5}$) with a white dwarf companion. Because it is a
bright pulsar for which high precision timing measurements can be made
and because the orbital inclination is high, measurements have been
made of the PK parameters $r$ and $s$.

We have reanalyzed all of the available data, extending from January
1986 to January 1994.  (For the data collected after mid-1989, taken
with the Princeton Mark~III timing system (\cite{skn+92}), a new
algorithm was used for identifying times and frequencies where the
signal strength was enhanced by interstellar scintillation and
weighting this data in subsequent analysis.)  The details and complete
timing solution will be published elsewhere.  The Shapiro parameters are
$r=Gm_2/c^3=0.248(11)M_\odot$ and $s=\sin i=0.9993(2)$.  The resulting
limits on the pulsar mass are $m_1=1.41(10)M_\odot$ (68\% confidence;
the 95\% confidence region will be about twice as large). The new
measurement is in good agreement with previous values:
$m_1=1.27^{+0.23}_{-0.15}M_\odot$ (\cite{rt91a}) and
$m_1=1.50^{+0.26}_{-0.14}$ (\cite{ktr94}).
\footnote{Note that both Ryba
  and Taylor (1991)\nocite{rt91a} and Kaspi \etal (1994)\nocite{ktr94}
  incorrectly (over)estimate the error on $m_1$ by calculating the
  {\it joint} 68\% confidence bound on $r=T_\odot m_2$ and $s=\sin i$
  and then interpreting this as a 68\% confidence bound on each of
  $m_2$ and $\sin i$.}

The $P_b$--$m_2$ relation predicts $0.19<m_2<0.26 M_\odot$, in good
agreement with the timing measurement.  The corresponding upper limit
to the pulsar mass is $m_1<1.51M_\odot$.  As noted above,
scintillation measurements have also been used to limits $\sin
i\geq0.94$ (\cite{dcww88}).

\subsubsection{PSR~J2019+2425}

PSR~J2019+2425 is in a 76.5~day orbit.  The $P_b$--$m_2$ relation
gives a limit on the companion mass $0.264 <m_2< 0.354M_\odot$; the
resulting upper limit on the pulsar mass is $m_1<1.68M_\odot$.

\subsection{Neutron star/main sequence binaries}
\label{sec:nsms}

\subsubsection{PSR~J0045$-$7319}
\label{sec:0045}

The pulsar J0045$-$7319 is the only known pulsar in the Small
Magellanic Cloud. It is in a binary orbit, with mass function
$f=2.17M_\odot$ (\cite{kjb+94}). The companion has been identified as
a B1~V star.  The radial velocity of the companion has been measured,
giving a mass ratio $q=m_2/m_1=6.3\pm1.2$ (\cite{bbs+95}) We have
compared the observed optical luminosity $L=1.2\times10^4L_\odot$ and
temperature $T_{\rm eff}=2.4(1)\times10^4$~K to the grids of stellar
models calculated by Schaller \etal (1992)\nocite{ssmm92} for the low
metallicity ($Z=0.001$) appropriate for the SMC, and estimate the
companion mass to be $10\pm1M_\odot$. The pulsar mass is then
$m_1=m_2/q=1.58\pm0.34M_\odot$ (68\% confidence), where the
uncertainty is dominated by the uncertainty in the amplitude of the
companion's radial velocity curve.

\section{Discussion}
\label{sec:stat}

For a dozen neutron stars, useful mass constraints are available with
no assumptions beyond the applicability of the general relativistic
equations of orbital motion to binary pulsar systems.  Ten of these
stars are members of double neutron star binaries.  With the possible
exception of PSR~B2127+11C, in the globular cluster M15, the pulsar in
each system is believed to have undergone a short period of mass
accretion during a high-mass X-ray binary phase ($\Delta
m\sim10^{-3}M_\odot$, Taam and van den Heuvel 1986).  The companion
stars have not undergone accretion; their masses most directly
preserve information about the initial mass function of neutron stars.

Only two ``millisecond'' pulsars, the end products of extended
mass transfer in low-mass X-ray binaries, have interesting mass
estimates based on GR alone: PSRs~B1802$-$07 and B1855+09.  Because
such pulsars must accrete $\sim0.1M_\odot$ to reach millisecond periods
(\cite{tv86}), and much more ($\sim0.7M_\odot$) in some field decay
models (e.g., \cite{vb95a}), obtaining additional mass
measurements of millisecond pulsars is of particular interest in
testing evolutionary models and in locating the maximum neutron star
mass.

As noted in \S\ref{sec:cmrr}, the $P_b$--$m_2$ relation can be used to
estimate the companion mass in recycled binary systems with circular
orbits and orbital periods $P_b\gtrsim 3$~d.  There are now thirteen
such millisecond ($P_{\rm spin}<10$~ms) pulsars known, excluding those
in globular clusters (where gravitational interactions may have
significantly perturbed the orbital parameters since spin-up).  In
each case, the measured mass function and the inferred companion mass,
together with the requirement that $\sin i<1$, then yields an upper
limit on the mass of the pulsar itself.  A number of systems in which
this upper limit is particularly constraining have been mentioned in
\S\ref{sec:nswd}.

Additional constraints on the neutron star mass in these
systems can be derived using statistical arguments, given a prior
assumption about the distribution of binary inclinations.  The
simplest such assumption is that the binaries are randomly oriented on
the sky, though biases toward high or low inclinations are possible
in some models (\S\ref{sec:uniform}).  
However, as discussed below, we
believe there is currently no evidence for such a bias, so for the
remainder of this discussion we assume random orbital orientations.

\begin{figure}
\plotone{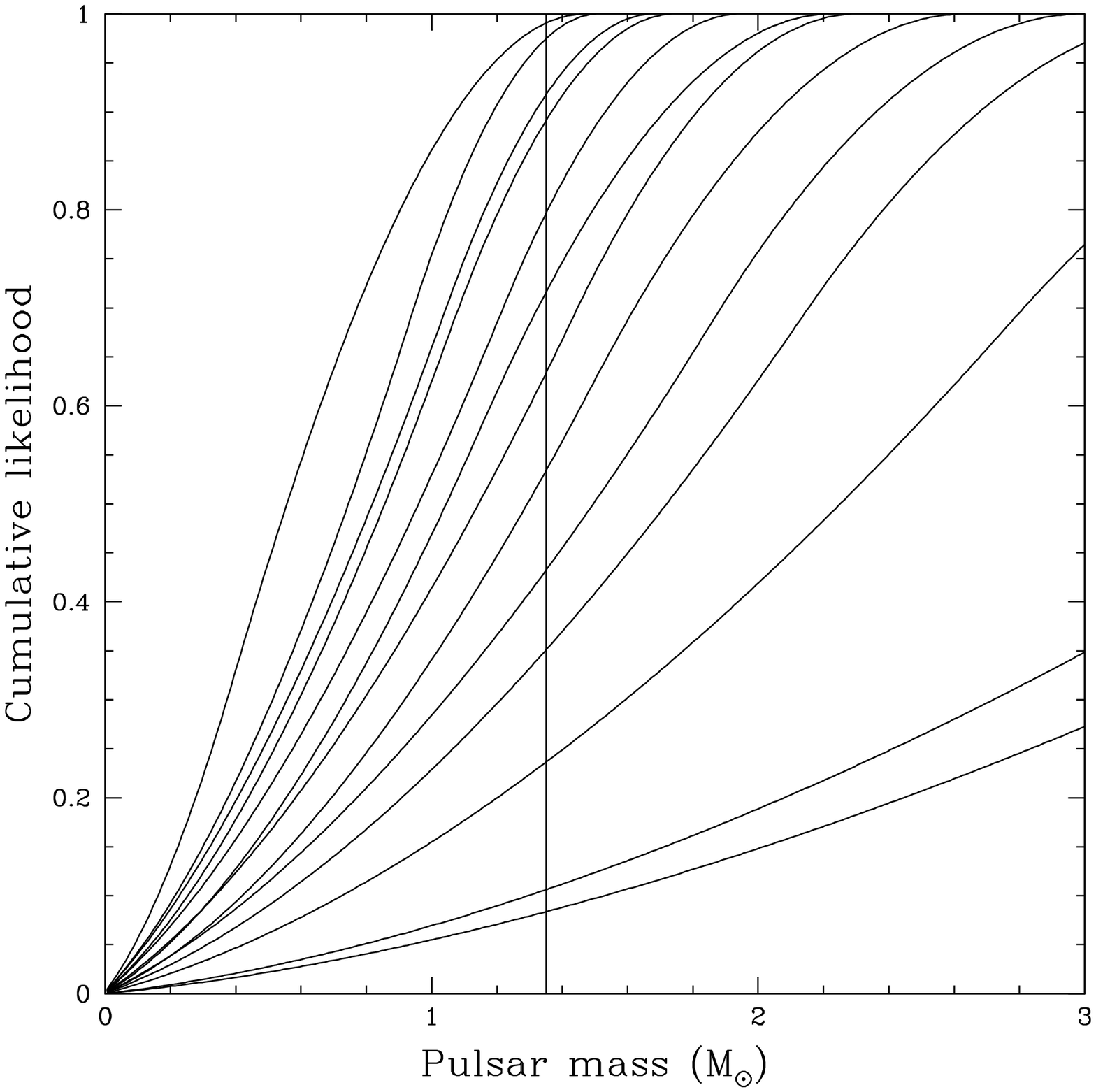}
\figcaption{Cumulative probabilities $\int\,dm_1\,p(m_1;f, P_b)$ for
  the 13 pulsars in Table~\ref{tab:pbm2}, as described in the text. A
  vertical line is shown at $1.35M_\odot$.
\label{fig:cumprob}}
\end{figure}

For an individual system, we are interested in the probability
distribution\footnote{We adopt the notation that $p(x ; A)$ is the
(marginal) probability density for the random variable $x$, where $x$
depends upon the parameter $A$.  Also, $p(x|y; A)$ is the conditional
probability density for the random variable $x$ for a given value of
the random variable $y$ and parameter $A$.}  $p(m_1; f, P_b)$ for the
neutron star mass $m_1$ given the measured mass function $f$ and
binary period $P_b$.  We can neglect the measurement uncertainty in
$f$ and $P_b$.  Then, the probability distribution for $m_1$ can be
written schematically as
\begin{equation}
p(m_1; f, P_b) = \int_0^1 d(\cos i) 
        \int_{m_{2,{\rm min}}(P_b)}^{m_{2,{\rm max}}(P_b)} dm_2\,
        p(m_2; P_b)\, p(\cos i)\, p(m_1|m_2,\cos i; f)  ,
\end{equation}
where $m_1=f^{-1/2}(m_2 \sin i)^{3/2}-m_2$ is restricted to positive
values.  We have evaluated this numerically for each system, assuming
that $p(\cos i)$ is uniform between zero and unity and that $p(m_2;
P_b)$ is uniformly distributed within the appropriate factor (see
\S\ref{sec:cmrr}) of the $m_2$ implied by equations (9)--(11).  Not
surprisingly, the width of the distribution $p(m_1; f, P_b)$ is
dominated by the range of allowed $\cos i$ rather than the uncertainty
in $m_2$ for a given $P_b$.  For each of the 13 binaries, we have
plotted the cumulative distribution CDF$(m_1)=\int_0^{m_1} p(m_1')\,
dm_1'$ in Figure~\ref{fig:cumprob}.\placefigure{fig:cumprob}

The median and 68\% and 95\% confidence regions for each pulsar mass
is given in Table~\ref{tab:pbm2}.  \placetable{tab:pbm2} Although
several of the pulsars have, under the assumptions made, most likely
masses well above $2M_\odot$, some such results are expected even if
all the masses are quite low.  In fact, in only one case of the 13
pulsars does $1.35M_\odot$ lie outside the 95\% central confidence
region (J1045$-$4509), and in 6 cases of 13 is $1.35M_\odot$ excluded
at 68\% confidence, consistent with chance.

\begin{figure}
\plotone{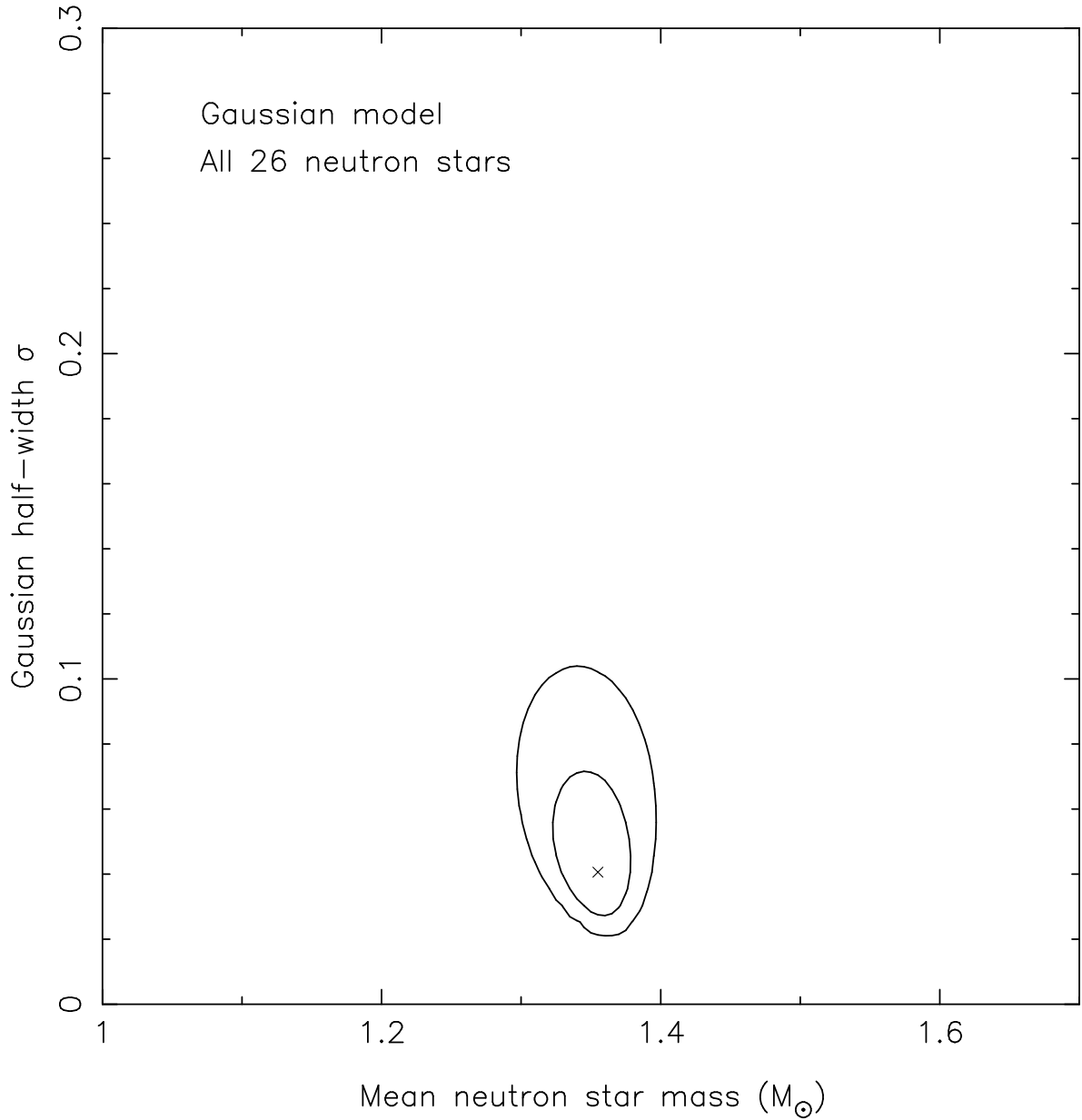}
\figcaption{Maximum likelihood estimate of the mean and standard
  deviation, $\hat{m}$ and $\sigma$, of a gaussian neutron star mass
  distribution. 
  The maximum likelihood solution is marked with a cross, and contours 
  indicate 68\% and 95\% confidence regions.
\label{fig:all26}}
\end{figure}

\begin{figure}
\plotone{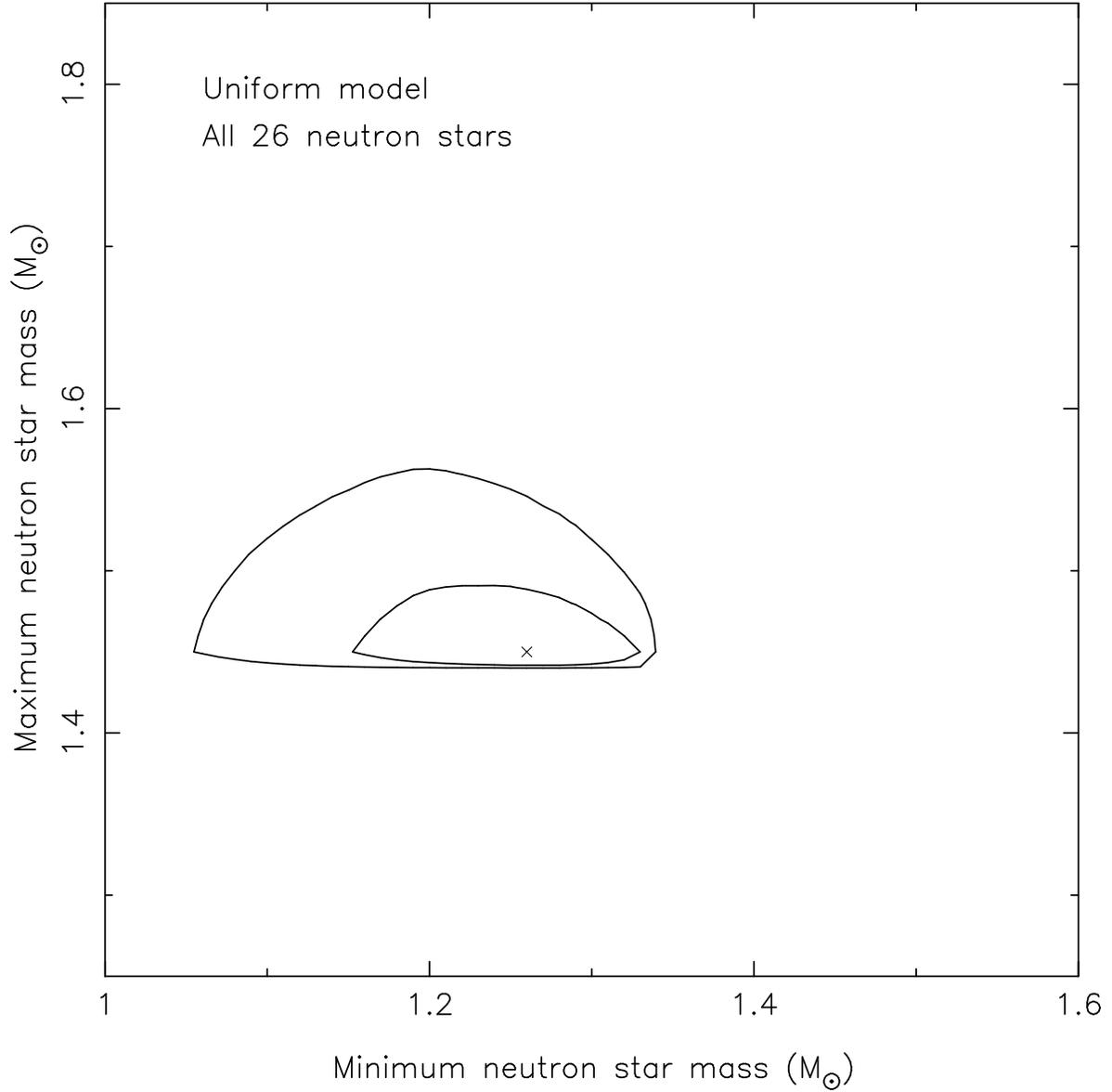}
\figcaption{Maximum likelihood estimate of the minimum and maximum
  neutron star mass, $m_l$ and $m_u$, assuming masses are uniformly
  distributed between the upper and lower bounds.
  The maximum likelihood solution is marked with a cross, and contours 
  indicate 68\% and 95\% confidence regions.
\label{fig:all26uni}}
\end{figure}

It is interesting to ask whether a single, simple distribution of
neutron star masses is consistent with all of our observational
constraints.  We considered two models for this question: a
Gaussian distribution of masses with mean $\hat{m}$ and standard
deviation $\sigma$, and a uniform distribution of masses between $m_l$
and $m_u$ ({\it cf.} Finn 1994).  A maximum likelihood analysis was used
to estimate the parameters $\hat{m}$, $\sigma$, $m_l$, and $m_u$
(assuming a uniform prior distribution for all four parameters).  The
resulting 68\% and 95\% joint confidence limits on $\hat{m}$ and
$\sigma$ are shown in Figure~\ref{fig:all26}, and on $m_l$ and $m_h$
in Figure~\ref{fig:all26uni}.  \placefigure{fig:all26}
\placefigure{fig:all26uni} In each model, the distribution of
neutron star masses is remarkably narrow: the maximum likelihood
solutions are $\hat{m}=1.35M_\odot$ and $\sigma=0.04M_\odot$, and
$m_l=1.26M_\odot$ and $m_u=1.45M_\odot$.

Of course, {\em any} model (even a poor one) will yield maximum
likelihood parameters for a given data set.  However, it is obvious by
inspection that both the Gaussian and uniform distributions for the
neutron star mass are good fits to the extremely narrow observed range
of neutron star masses in the double neutron star binaries.  While it
is difficult to quantify the goodness-of-fit for the entire data set,
because of the diverse assumptions made in the various mass estimates
and the sometimes highly non-gaussian error estimates, we can easily
test some neutron star subsamples against the maximum likelihood
gaussian model $m_1=1.35\pm0.04M_\odot$.  For the thirteen
neutron-star--white-dwarf binaries, we used a Monte Carlo technique to
evaluate the fit quality.  For each binary (with its measured $P_b$
and $f$), we simulated a large number of Monte Carlo trials where the
neutron star mass $m_1$ was drawn from the maximum likelihood model,
$m_2$ was drawn from the appropriate uniform distribution implied by
$P_b$, and $\cos i$ was drawn from a uniform distribution.  The Monte
Carlo trials were then used to construct the probability distribution
for the mass function, and this distribution was used to compute the
cumulative probability for the measured mass function,
$p(f^\prime<f)$.  If the model and the associated assumptions are
correct, then the cumulative probabilities for the 13 measured mass
functions should be consistent with a uniform distribution betwen zero
and unity ({\it cf.} a classical $V/V_{\rm max}$ test, Schmidt 1968). A
KS test of the distribution 
shows consistency with a uniform distribution at the
improbably good 99\% level (Figure~\ref{fig:ks}). \placefigure{fig:ks}

For the ten stars for which gaussian error estimates $\sigma_e$ are
available (both stars in the relativistic binaries B1534+12, B1913+16,
and B2127+11C, as well as J1012+5307, J1713+0747, B1855+09, and
J0045$-$7319), we can calculate a $\chi^2$ statistic,
$\sum(m-\hat{m})^2/(\sigma^2+\sigma_e^2)=7.5$, consistent with
expectations for a chi-square distribution with $10-2$ degrees of freedom.

We conclude, therefore, that at least in the radio pulsar systems,
there is no evidence for neutron star masses above about
$1.45M_\odot$.  Indeed, the data appear very well modeled by very narrow
distributions centered around $1.35M_\odot$.


\section{Summary}
\label{sec:summ}

\begin{figure}
\plotone{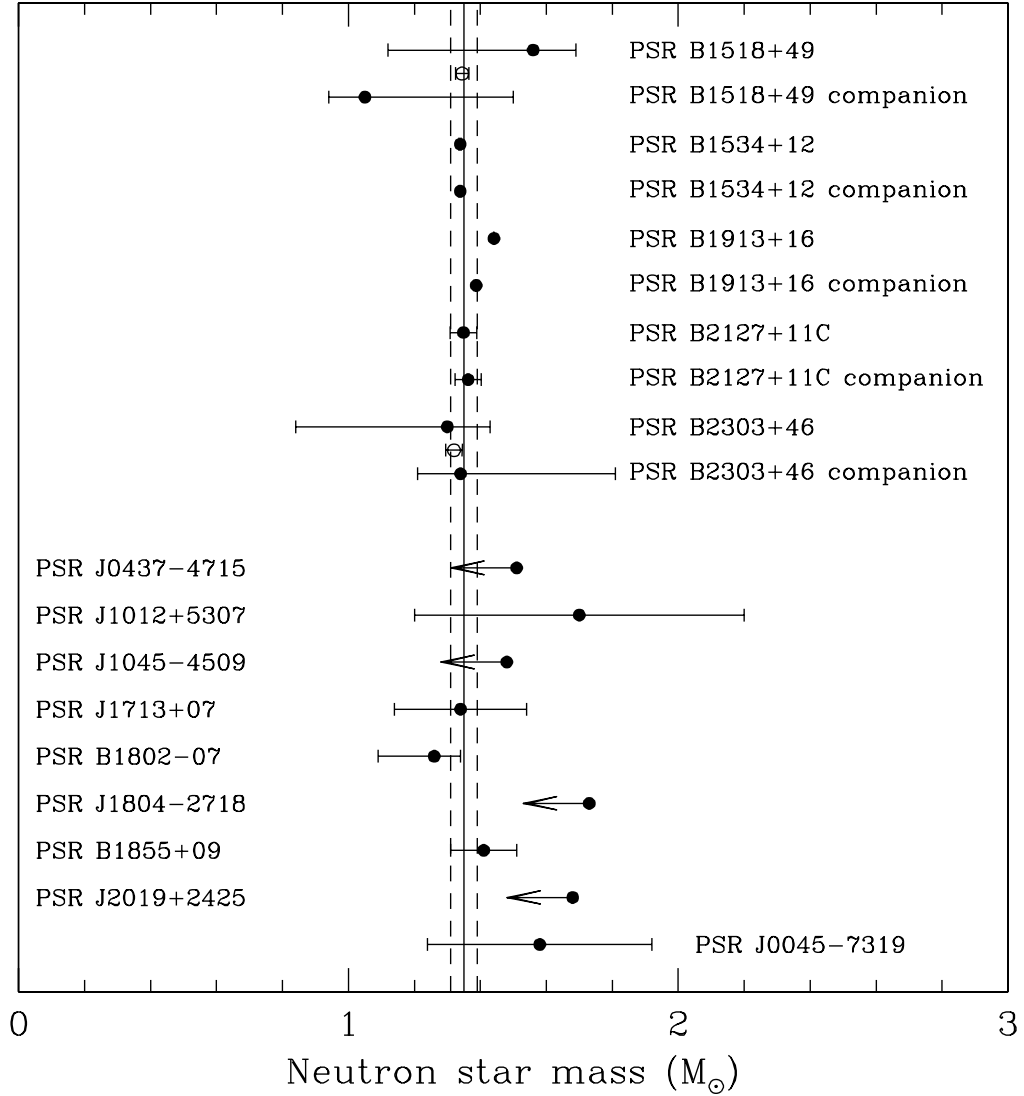} 
\figcaption{Neutron star masses from
    observations of radio pulsar systems. All error bars indicate
    central 68\% confidence limits, except upper limits are one-sided
    95\% confidence limits. Five double neutron star systems
    are shown at the top of the diagram.  In two cases, the average
    neutron star mass in a system is known with much better accuracy
    than the individual masses; these average masses are indicated
    with open circles.  Eight neutron-star--white-dwarf binaries are
    shown in the center of the diagram, and one
    neutron-star--main-sequence-star binary is shown at
    bottom. Vertical lines are drawn at $m=1.35\pm0.04M_\odot$.
\label{fig:massrev}}
\end{figure}

There are now 26 neutron stars in binary radio pulsar systems for
which useful mass constraints can be derived.  Of these, about half
are neutron star-white dwarf binaries in which the mass determination
depends on the validity of the $P_b$--$m_2$ relation and the isotropy
of the binary orbits with respect to the line-of-sight, as discussed
in the previous section.  All other mass constraints are listed in
Table~\ref{tab:masssum} and shown in Figure~\ref{fig:massrev}.
\placetable{tab:masssum} \placefigure{fig:massrev}

Although we defer a full discussion of the underlying neutron star mass
distribution and it implications for neutron star formation and
evolution until after analysis of the X-ray binary systems (Paper~II),
we note here a few points of particular interest about the radio
pulsar binaries.  Figure~\ref{fig:massrev} is striking primarily for the
very small variations in the masses of well measured stars.  This has,
of course, been noted before (e.g., Thorsett \etal
1993)\nocite{tamt93}, but it remains surprising that no new mass
measurements differ greatly from $1.4M_\odot$. In the five double neutron
star binaries for which a relativistic periastron advance yields an
accurate total (and hence average) mass, the average neutron star
masses vary by less than 7\%.

The most surprising implication of the current results is that there
is little evidence for mass tranfer of $0.1M_\odot$ or more in the
millisecond pulsar systems.  It is important, then, to reiterate the
assumptions upon which this conclusion rests: (1) the $P_b$--$m_2$
is correct, at least within the (modest) claimed precision, and (2)
binary orbits are randomly oriented with respect to the line of
sight. There are good prospects for testing both assumptions.

The reliability of the $P_b$--$m_2$ relation depends principally upon the
core-mass--radius relation for red giants (eqns.~\ref{eqn:rapp1} and
\ref{eqn:rapp2}).  The latter relation can be tested observationally
by careful study of nearby red giants.  Precise measurements of
bolometric flux and angular size (through optical/infrared photometry
and interferometry; see, for example, Dyck \etal 1996, Perrin \etal
1998) together with accurate distance measurements (e.g., using {\em
  Hipparcos}) can be used to probe the relationship between luminosity
and radius.  Since the core-mass--luminosity relation for red giants
should be inherently more precise than the core-mass--radius relation
(Rappaport \etal 1995 and references therein), such observations would
provide an effective test of the core-mass--radius (and hence
$P_b$--$m_2$) relation.  Further, we hope that the power of the
$P_b$--$m_2$ relation as a statistical tool will encourage more
detailed theoretical investigations of, in particular, the extension
of the relation to orbital periods below about three days, where X-ray
heating and bloating of the companion star become important.

Although any technique that limits $\sin i$ (\S\ref{sec:sini}; e.g.,
polarization or scintillation studies) can be used to test the
hypothesis that orbits are randomly inclined, the most precise
measurements will come from pulsar timing and Shapiro time delay
measurements.  In Table~\ref{tab:pbm2} we list the mean predicted amplitude
of the Shapiro delay signal $\Delta t_S=2m_2T_\odot\log(1-\sin i)$ for
each of the thirteen pulsars discussed in \S\ref{sec:stat}, under the
assumption that the neutron stars are all $1.4M_\odot$ and the
companion masses are the central value predicted by the $P_b$--$m_2$
relation.  Of these systems, only B1855+09 has a well measured signal
(\cite{rt91a,ktr94}); it and the less studied
J2019+2425 have the largest predicted signals.  (It is also interesting
to note that of the pulsars in Table~\ref{tab:pbm2}, the only two that
have been clearly shown to have ``classical'' interpulse emission
separated from the main radio emission peak by $\sim180^\circ$ are
B1855+09 and J1804$-$2718.  These are two of the four pulsars that
the $P_b$--$m_2$ relation suggests are observed on lines of sight that
are most nearly equatorial, as would be expected for systems in which
both magnetic poles are seen.)

\begin{figure}
\plotone{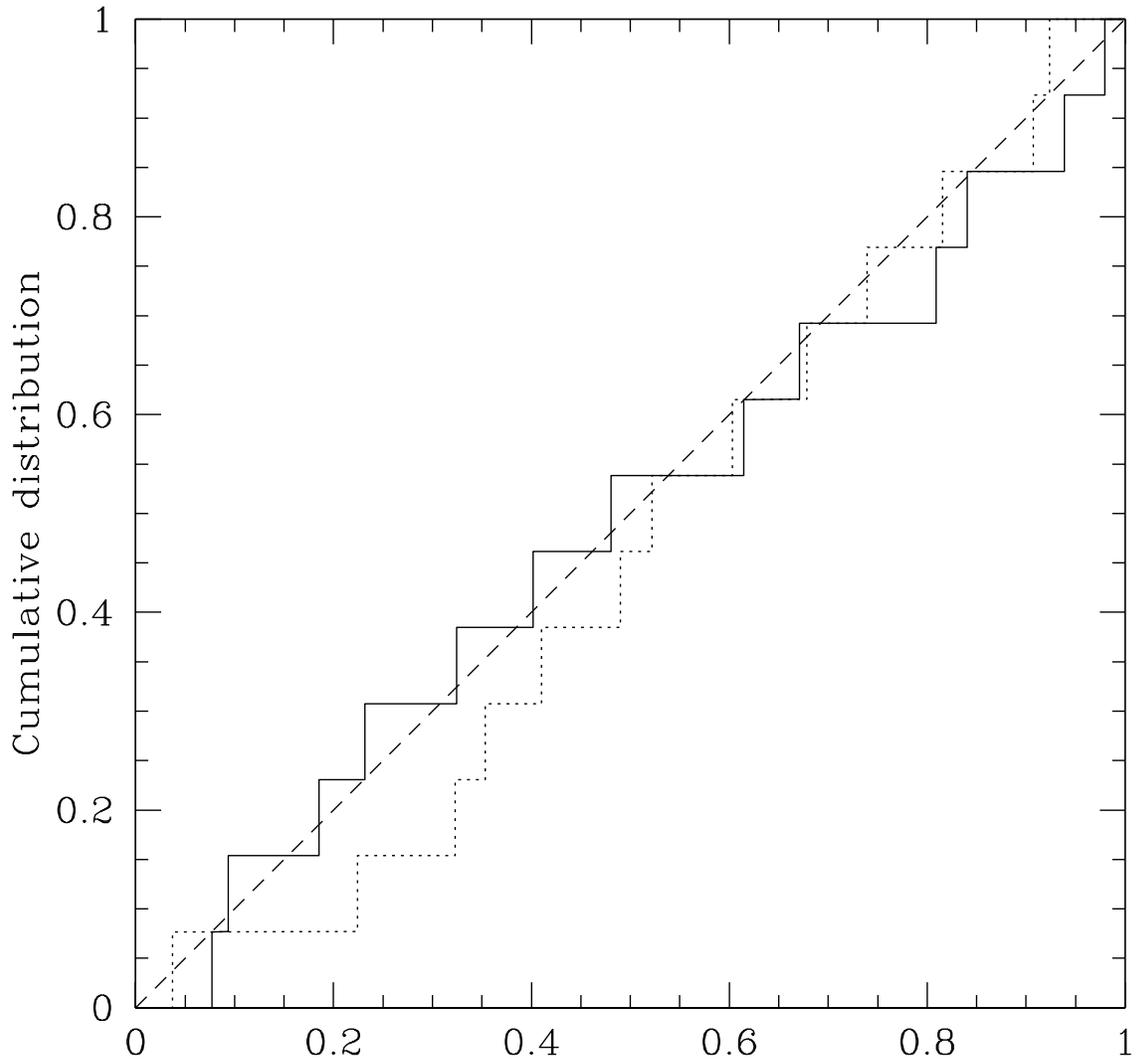}
\figcaption{Cumulative distribution of $p(f^\prime<f)$ (solid line,
  \S\ref{sec:stat}) and
of $\cos i$ (dotted line, \S\ref{sec:summ}) for the thirteen millisecond
pulsars with $P_b>3$~days and $P<10$~ms, assuming a pulsar mass
distribution $m_1=1.35\pm0.04M_\odot$ and companion mass distribution
as predicted by the $P_b$--$m_2$ relation (see text). Each
distribution is consistent with uniform (dashed line, 
KS probability of 99\% and
81\%, respectively), as expected if orbital inclinations are randomly
distributed and the $P_b$--$m_2$ relation correctly predicts the
companion mass distribution.
\label{fig:ks}}
\end{figure}

An assumed pulsar mass distribution and the $P_b$--$m_2$
relation allow us to test our assumption that the orbital inclinations 
are randomly distributed: i.e., $\cos i$ is uniformly distributed. In
Figure~\ref{fig:ks} we show the cumulative distribution of $\cos i$
for the thirteen pulsar--white-dwarf binaries discussed above, using
$\sin i$ values from Table~\ref{tab:pbm2} except for B1855+09, for
which a better estimate is available from timing. We find that the
measured values are consistent with uniform at the 
81\% level. Because the pulsar spin and orbital angular momenta are
most likely aligned, we further conclude that there is no evidence for 
either alignment or counteralignment of pulsar magnetic fields in
millisecond pulsars, contrary to some predictions (e.g., \cite{rud91}).


\acknowledgments The observations on which this paper is based were
carried out by many people.  SET particularly thanks his collaborators
Z. Arzoumanian, D. J. Nice, and J. H. Taylor.  V. M. Kaspi, L. Rawley,
and M. Ryba all contributed significantly to the PSR~B1855+09 dataset,
as did A. Vasquez and other staff members of the Arecibo Observatory,
operated by Cornell University for the U. S. National Science
Foundation.  Observations were also made with the facilities of the
National Radio Astronomy Observatory.  We thank M.~H. van~Kerkwijk for
sharing results in advance of publication, and D. C. Backer for
valuable discussions. The research of SET is
supported by the NSF.  DC is supported by a NASA Compton GRO
Postdoctoral Fellowship, under grant NAG 5-3109.

\clearpage

\begin{deluxetable}{llllc}
\small
\tablecolumns{5}
\tablewidth{0pc}
\tablecaption{Binary radio pulsar systems\label{tab:allbins}\tablenotemark{a}}
\tablehead{
\colhead{PSR name} & \colhead{$P_{\rm spin}$ (s)} & \colhead{$P_b$ (d)} &
  \colhead{$e$} & \colhead{Notes} }
\startdata
\cutinhead{\em Double neutron star binaries}
J1518+4904&0.040935&\phn\phn\phn 8.634&0.24948&\nl
B1534+12&0.037904&\phn\phn\phn 0.421&0.27368&\nl
B1913+16&0.05903&\phn\phn\phn 0.323&0.61713&\nl
B2127+11C&0.030529&\phn\phn\phn 0.335&0.68141&1\nl
B2303+46&1.066371&\phn\phn 12.34&0.65837&\nl
\cutinhead{\em Neutron star/white dwarf binaries}
B0021$-$72E&0.003536&\phn\phn\phn 2.257&0.000&1\nl
B0021$-$72I&0.003485&\phn\phn\phn 0.226&0.00&1\nl
B0021$-$72J&0.002101&\phn\phn\phn 0.121&0.00&1\nl
J0034$-$0534&0.001877&\phn\phn\phn 1.589&0.0000&\nl
J0218+4232&0.002323&\phn\phn\phn 2.029&0.00000&\nl
J0437$-$4715&0.005757&\phn\phn\phn 5.741&0.00000&\nl
J0613$-$0200&0.003062&\phn\phn\phn 1.199&0.00000&\nl
B0655+64&0.195671&\phn\phn\phn 1.029&0.00000&\nl
J0751+1807&0.003479&\phn\phn\phn 0.263&0.0000&\nl
B0820+02&0.864873& 1232.47&0.01187&\nl
J1012+5307&0.005256&\phn\phn\phn 0.605&0.00000&\nl
J1022+10&0.016453&\phn\phn\phn 7.805&0.0001&\nl
J1045$-$4509&0.007474&\phn\phn\phn 4.084&0.00000&\nl
B1310+18&0.033163&\phn 255.8&0.002&1\nl
J1455$-$3330&0.007987&\phn\phn 76.175&0.00017&\nl
J1603$-$7202 & 0.014842 & \phn\phn\phn6.309 & 0.0000 & 2\nl
B1620$-$26&0.011076&\phn 191.443&0.02531&1,4\nl
J1640+2224&0.003163&\phn 175.461&0.0008&\nl
B1639+36B&0.003528&\phn\phn\phn 1.259&0.005&1\nl
J1643$-$1224&0.004622&\phn 147.017&0.00051&\nl
J1713+0747&0.004570&\phn\phn 67.825&0.00007&\nl
B1718$-$19&1.004037&\phn\phn\phn 0.258&0.000&1\nl
B1744$-$24A&0.011563&\phn\phn\phn 0.076&0.0000&1\nl
B1800$-$27&0.334415&\phn 406.781&0.00051&\nl
B1802$-$07&0.023101&\phn\phn\phn 2.617&0.212&1\nl
J1804$-$2717 & 0.009343 & \phn\phn11.129 & 0.00004 & 2\nl
B1820$-$11&0.279828&\phn 357.762&0.79462&3\nl
B1831$-$00&0.520954&\phn\phn\phn 1.811&0.000&\nl
B1855+09&0.005362&\phn\phn 12.327&0.00002&\nl
J1910+0004&0.003619&\phn\phn\phn 0.141&0.00&1\nl
J1911$-$1114 & 0.003626 & \phn\phn\phn2.717 & 0.0000 & 2\nl
B1953+29&0.006133&\phn 117.349&0.00033&\nl
B1957+20&0.001607&\phn\phn\phn 0.382&0.00000&\nl
J2019+2425&0.003935&\phn\phn 76.512&0.00011&\nl
J2033+17&0.005949&\phn\phn 56.2&0.00&\nl
J2129$-$5721 & 0.003726 & \phn\phn\phn6.625 & 0.0000 & 2\nl
J2145$-$0750&0.016052&\phn\phn\phn 6.839&0.00002&\nl
J2229+2643&0.002978&\phn\phn 93.016&0.00026&\nl
J2317+1439&0.003445&\phn\phn\phn 2.459&0.00000&\nl
\cutinhead{\em Neutron star/main sequence binaries}
J0045$-$7319&0.926276&\phn\phn 51.169&0.808&\nl
B1259$-$63&0.047762& 1236.724&0.86993&\nl
\cutinhead{\em Pulsar with planetary mass companions}
B1257+12&0.006219&\phn\phn 66.536&0.0182&\nl
        & 2nd planet  &\phn\phn 98.223&0.0264&\nl
\enddata
\tablenotetext{a}{Unless otherwise indicated, all data are from the 1995 
revision of the Princeton Pulsar Catalog (Taylor {\it et al.}
1995).\nocite{tmlc95}} 
\tablecomments{(1) Believed to be a globular cluster member; (2)
Lorimer {\it et al.} 1996\nocite{llb+96}; (3) may be a double neutron
star system; (4) probably a triple system (\cite{tat93}).}
\end{deluxetable}
\newpage

\begin{deluxetable}{lllclllllll}
\footnotesize
\tablecolumns{11}
\tablewidth{7in}
\tablecaption{Mass estimates from the $P_b$--$m_2$
  relation.\label{tab:pbm2}}
\tablehead{
&  &  &
& \multicolumn{5}{c}{Pulsar mass\tablenotemark{b}\ ($M_\odot$)} &
\multicolumn{2}{c}{Predicted:\tablenotemark{c}}\nl \cline{5-9} \cline{10-11}
& \colhead{$P_b$} & \colhead{$f$} & \colhead{$m_2$\tablenotemark{a}} &
\colhead{95\%} & \colhead{68\%} & & \colhead{68\%} &
\colhead{95\%} & & \colhead{$\Delta t_S$}\nl
\colhead{Pulsar} & \colhead{(days)} & \colhead{($10^{-3}M_\odot$)} &
\colhead{($M_\odot$)} & \colhead{lower} & \colhead{lower} & \colhead{median}&
\colhead{upper} & \colhead{upper}&\colhead{$\sin i$}&\colhead{($\mu$s)}
}
\startdata
J0437$-$4715 & \phn\phn5.741 & \phn1.243 & 0.164
& 0.106 & 0.472 & 1.044 & 1.573 & 1.971& 0.87 & \phn3.5 \nl
J1045$-$4509 & \phn\phn4.083 & \phn1.765 & 0.132
& 0.047 & 0.234 & 0.557 & 0.965 & 1.273& 0.97 & \phn6.2 \nl
J1455$-$3330 & \phn76.174 & \phn6.272 & 0.305
& 0.098 & 0.491 & 1.144 & 1.681 & 2.058& 0.85 & \phn5.9 \nl
J1640+2224   & 175.460 & \phn5.907 & 0.351
& 0.135 & 0.651 & 1.495 & 2.190 & 2.681 & 0.73 & \phn4.6 \nl
J1643$-$1224 & 147.017 & \phn0.783 & 0.341
& 0.553 & 2.106 & 4.439 & 6.335 & 7.674 & 0.38 & \phn1.6 \nl
J1713+0747   & \phn67.825 & \phn7.896 & 0.299
& 0.078 & 0.405 & 0.960 & 1.419 & 1.741 & 0.91 & \phn7.6 \nl
J1804$-$2718 & \phn11.128 & \phn4.137 & 0.212
& 0.073 & 0.369 & 0.856 & 1.265 & 1.558 & 0.94 & \phn6.7 \nl
B1855+09     & \phn12.327 & \phn 5.557 & 0.219
& 0.060 & 0.315 & 0.745 & 1.101 & 1.351 & 0.97 & \phn9.0 \nl
B1953+29     & 117.349 & \phn2.417 & 0.328
& 0.236 & 1.024 & 2.251 & 3.251 & 3.955 & 0.58 & \phn2.8 \nl
J2019+2425   & \phn76.511 & 10.686 & 0.305
& 0.062 & 0.338 & 0.818 & 1.218 & 1.500 & 0.95 & 10.0\nl
J2033+17     & \phn56.2 & \phn2.75 & 0.290
& 0.172 & 0.772 & 1.720 & 2.489 & 3.029 & 0.67 & \phn3.2 \nl
J2129$-$5721 & \phn\phn6.625 & \phn1.049 & 0.176
& 0.138 & 0.591 & 1.293 & 1.900 & 2.348 & 0.80 & \phn2.7 \nl
J2229+2643   & \phn93.015 &\phn0.839 & 0.315
& 0.463 & 1.789 & 3.787 & 5.409 & 6.552 & 0.42 & \phn1.7 \nl
\enddata
\tablenotetext{a}{Central value from $P_b$--$m_2$ relation,
  eqn.~\ref{eqn:pb} and eqns.~\ref{eqn:rapp1} and \ref{eqn:rapp2}.}
\tablenotetext{b}{Median and central 68\% and 95\% confidence bounds.}
\tablenotetext{c}{Mean value of $\sin i $ and Shapiro delay amplitude
  $\Delta t_S=2m_2T_\odot\log(1-\sin i)$, 
  assuming a gaussian underlying neutron
  star mass distribution $m_1=1.35\pm0.04$ and uniform $m_2$ in the
  range allowed by the $P_b$--$m_2$ relation (\S\ref{sec:summ}).}
\end{deluxetable}
\clearpage

\begin{deluxetable}{lllll}
\scriptsize
\tablecolumns{5}
\tablewidth{0pt}
\tablecaption{Radio pulsar mass summary\label{tab:masssum}}
\tablehead{
 & \colhead{Median} & \colhead{68\%} & \colhead{95\%} \nl
\colhead{Star}& \colhead{mass ($M_\odot$)} & \colhead{central imits} &
\colhead{central limits} &\colhead{Notes\tablenotemark{a}}
}
\startdata
\cutinhead{\em Double neutron star binaries}
\sidehead{J1518+4904}
pulsar & 1.56 & $+0.13/-0.44$ & $+0.20/-1.20$ & GR, RO\nl
companion & 1.05&$+0.45/-0.11$&$+1.21/-0.14$&GR, RO\nl
average & 1.31 & $\pm0.035$ & $\pm0.07$ & GR\nl
\sidehead{B1534+12}
pulsar & 1.339 & $\pm 0.003$ & $\pm 0.006$ & GR\nl
companion & 1.339 & $\pm 0.003$ & $\pm 0.006$ & GR\nl
\sidehead{B1913+16}
pulsar & 1.4411 & $\pm 0.00035$ & $\pm 0.0007$ & GR \nl
companion &1.3874 & $\pm 0.00035$ & $\pm 0.0007$& GR \nl
\sidehead{B2127+11C}
pulsar& 1.349 & $\pm 0.040$ & $\pm 0.080$ & GR\nl
companion & 1.363 & $\pm 0.040$ & $\pm 0.080$ & GR\nl
average & 1.3561 & $\pm 0.0003$ & $\pm 0.0006$ & GR\nl
\sidehead{B2303+46}
pulsar & 1.30 & $+0.13/-0.46$ & $+0.18/-1.08$ & GR, RO\nl
companion & 1.34 & $+0.47/-0.13$ &$+1.08/-0.15$&GR, RO\nl
average & 1.32 & $\pm 0.025$ & $\pm 0.05$ & GR\nl
\cutinhead{\em Neutron star/white dwarf binaries}
J0437$-$4715 &\nodata &\nodata & $<1.51$ & $\dot x$, $P_bm_2$ \nl
J1012+5307 & 1.7 & $\pm 0.5$ & $\pm 1.0$ & Opt \nl
J1045$-$4509 &\nodata &\nodata & $<1.48$ & $P_bm_2$ \nl
J1713+0747 & 1.45 & $\pm 0.31$ & $\pm 0.62$ & $P_bm_2$, GR\nl
           & 1.34 & $\pm 0.20$ & $\pm 0.40$ & $P_bm_2$, GR, Opt\nl
B1802$-$07 &\nodata  & $<1.39$ & $<1.45$ & GR\nl
           & 1.26 & $+0.08/-0.17$ & $+0.15/-0.67$ & GR, $P_bm_2$\nl
J1804$-$2718 &\nodata &\nodata & $<1.73$ & $P_bm_2$\nl
B1855+09 & 1.41 & $\pm 0.10$ & $\pm 0.20$ & GR \nl
J2019+2425 &\nodata &\nodata & $<1.68$ & $P_bm_2$\nl
\cutinhead{\em Neutron star/main sequence binaries}
J0045$-$7319 & 1.58 & $\pm 0.34$ & $\pm 0.68$ & Opt, MS\nl
\enddata
\tablenotetext{a}{Assumptions made in mass estimate: (GR) general
  relativistic binary model; (RO) random orbital orientation, or
  inclination angle uniform in $\cos i$; ($P_bm_2$) core mass orbital
  period relation; ($\dot x$) proper-motion-induced change in the
  projected semi-major axis; (Opt) optical companion observations;
  (MS) main sequence stellar model.
}
\end{deluxetable}
\clearpage


\newpage

\begin{thebibliography}{}

\bibitem[Arzoumanian 1995]{arz95}
Arzoumanian, Z. 1995, PhD thesis, Princeton University.

\bibitem[Baade and Zwicky 1934]{bz34b}
Baade, W. and Zwicky, F. 1934, Phys.\ Rev., 45, 138.

\bibitem[Backer 1998]{bac98}
Backer, D.~C. 1998, ApJ, 493, 873.

\bibitem[Bailyn 1993]{bai93}
Bailyn, C.~D. 1993, ApJ, 411, L83.

\bibitem[Becker and Tr\"umper 1993]{bt93a}
Becker, W. and Tr\"umper, J. 1993, Nature, 365, 528.

\bibitem[Bell, Bailes, and Bessell 1993]{bbb93}
Bell, J.~F., Bailes, M., and Bessell, M.~S. 1993, Nature, 364,
  603.

\bibitem[Bell \etal  1995]{bbs+95}
Bell, J.~F., Bessell, M.~S., Stappers, B.~W., Bailes, M., and Kaspi, V.~M.
  1995, ApJ, 447, L117.

\bibitem[Bergeron, Wesemael, and Fontaine 1991]{bwf91}
Bergeron, P., Wesemael, F., and Fontaine, G. 1991, ApJ, 367, 253.

\bibitem[Camilo 1995]{cam95a}
Camilo, F. 1995, PhD thesis, Princeton University.

\bibitem[Camilo, Foster, and Wolszczan 1994]{cfw94}
Camilo, F., Foster, R.~S., and Wolszczan, A. 1994, ApJ, 437, L39.

\bibitem[Chakrabarty and Thorsett 1998]{ct98}
Chakrabarty, D. and Thorsett, S.~E. 1998, in preparation.

\bibitem[Cordes 1986]{cor86}
Cordes, J.~M. 1986, ApJ, 311, 183.

\bibitem[Damour and Deruelle 1986]{dd86}
Damour, T. and Deruelle, N. 1986, Ann.\ Inst.\ H. Poincar\'e (Physique
  Th\'eorique), 44, 263.

\bibitem[Damour and Taylor 1992]{dt92}
Damour, T. and Taylor, J.~H. 1992, Phys.\ Rev.\ D, 45, 1840.

\bibitem[D'Antona and Mazzitelli 1990]{dm90}
D'Antona, F. and Mazzitelli, I. 1990, ARA\&A, 28, 139.

\bibitem[Danziger, Baade, and Della~Valle 1993]{dbd93}
Danziger, I.~J., Baade, D., and Della~Valle, M. 1993, AA, 276, 382.

\bibitem[Deich and Kulkarni 1996]{dk96}
Deich, W. T.~S. and Kulkarni, S.~R. 1996, in {\em {C}ompact {S}tars in
  {B}inaries: {IAU} Symposium 165}, ed.\ J. van Paradijs, E.~P.~J. van~del
  Heuvel, and E. Kuulkers, (Dordrecht: Kluwer), 279.

\bibitem[Dewey \etal  1988]{dcww88}
Dewey, R.~J., Cordes, J.~M., Wolszczan, A., and Weisberg, J.~M. 1988, in {\em
  Radio Wave Scattering in the Interstellar Medium, {AIP} Conference
  Proceedings {N}o. 174}, ed.\ J.M. Cordes, B.~J. Rickett, and D.~C. Backer,
  (New York: American Institute of Physics), 217.

\bibitem[Dyck \etal 1996]{dbvr96}
Dyck, H.~M., Benson, J.~A., van~Belle, G.~T., \& Ridgway, S.~T. 1996,
AJ, 111, 1705.

\bibitem[Eadie \etal 1971]{edj+71}
Eadie, W. T., Drijard, D., James, F. E., Roos, M., and Sadoulet,
B. 1971, {\em Statistical Methods in Experimental Physics},
(Amsterdam: North-Holland).

\bibitem[Finn 1994]{fin94}
Finn, L. S. 1994, Phys.\ Rev.\ Lett., 73, 1878.

\bibitem[Friedman \etal  1988]{fidp88}
Friedman, J.~L., Ipser, J.~R., Durisen, R.~H., and Parker, L. 1988,
  Nature, 336, 560.

\bibitem[Gil and Krawczyk 1997]{gk97}
Gil, J. and Krawczyk, A. 1997, MNRAS, 285, 561.

\bibitem[Halpern, Martin, and Marshall 1996]{hmm96}
Halpern, J.~P., Martin, C., and Marshall, H.~L. 1996, ApJ, 462, 908.

\bibitem[Hansen and Phinney 1998a]{hp98a}
Hansen, B. M.~S. and Phinney, E.~S. 1998a, MNRAS, 294, 557.

\bibitem[Hansen and Phinney 1998b]{hp98b}
Hansen, B. M.~S. and Phinney, E.~S. 1998b, MNRAS, 294, 569.

\bibitem[Hulse and Taylor 1975]{ht75a}
Hulse, R.~A. and Taylor, J.~H. 1975, ApJ, 195, L51.

\bibitem[Jenet \etal  1997]{jcpu97}
Jenet, F.~A., Cook, W.~R., Prince, T.~A., and Unwin, S.~C. 1997, 
PASP, 109, 707.

\bibitem[Johnston \etal  1993]{jlh+93}
Johnston, S. \etal  1993, Nature, 361, 613.

\bibitem[Johnston \etal  1992]{jml+92}
Johnston, S., Manchester, R.~N., Lyne, A.~G., Bailes, M., Kaspi, V.~M., Qiao,
  G., and D'Amico, N. 1992, ApJ, 387, L37.

\bibitem[Johnston, Walker, and Bailes 1996]{jwb96}
Johnston, S., Walker, M. A., and Bailes, M. 1996.
\newblock {\em Pulsars: Problems and Progress, {IAU} Colloquium 160}, San
  Francisco. Astronomical Society of the Pacific.

\bibitem[Jones and Lyne 1988]{jl88}
Jones, A.~W. and Lyne, A.~G. 1988, MNRAS, 232, 473.

\bibitem[Joss, Rappaport, and Lewis 1987]{jrl87}
Joss, P.~C., Rappaport, S., and Lewis, W. 1987, ApJ, 319, 180.

\bibitem[Kaspi \etal  1994]{kjb+94}
Kaspi, V.~M., Johnston, S., Bell, J.~F., Manchester, R.~N., Bailes, M.,
  Bessell, M., Lyne, A.~G., and D'Amico, N. 1994, ApJ, 423, L43.

\bibitem[Kaspi, Taylor, and Ryba 1994]{ktr94}
Kaspi, V.~M., Taylor, J.~H., and Ryba, M. 1994, ApJ, 428,
  713.

\bibitem[Lorimer \etal  1996]{llb+96}
Lorimer, D.~R., Lyne, A.~G., Bailes, M., Manchester, R.~N., D'Amico, N.,
  Stappers, B.~W., Johnston, S., and Camilo, F. 1996, MNRAS, 283, 1383.

\bibitem[Lundgren, Camilo, and Foster 1996]{lfc96}
Lundgren, S.~C., Camilo, F., and Foster, R.~S. 1996, In Johnston \etal
  1996, p.\ 497.

\bibitem[Lyne 1984]{lyn84}
Lyne, A.~G. 1984, Nature, 310, 300.

\bibitem[Manchester and Johnston 1995]{mj95}
Manchester, R.~N. and Johnston, S. 1995, ApJ, 441,
  L65.

\bibitem[Manchester and Taylor 1977]{mt77}
Manchester, R.~N. and Taylor, J.~H. 1977, {\em Pulsars}, (San Francisco:
  Freeman).

\bibitem[Navarro \etal  1997]{nms+97}
Navarro, J., Manchester, R.~N., Sandhu, J.~S., Kulkarni, S.~R., and Bailes, M.
  1997, ApJ, 486, 1019.

\bibitem[Nice, Sayer, and Taylor 1996]{nst96}
Nice, D.~J., Sayer, R.~W., and Taylor, J.~H. 1996, ApJ, 466, L87.

\bibitem[Oppenheimer and Volkoff 1939]{ov39}
Oppenheimer, J.~R. and Volkoff, G. 1939, Phys.\ Rev., 55, 374.

\bibitem[Perrin \etal 1998]{pcr+98}
Perrin, G., Coud\'{e}~du~Foresto, V., Ridgway, S.~T., Mariotti, J.-M.,
Traub, W.~A., Carleton, N.~P., \& Lacasse, M.~G. 1998, A\&A, 331,
619.

\bibitem[Podsiadlowski 1991]{pod91}
Podsiadlowski, P. 1991, Nature, 350, 136.

\bibitem[Rappaport and Joss 1997]{rj97}
Rappaport, S. and Joss, P.~C. 1997, ApJ, 486, 435.

\bibitem[Rappaport \etal  1995]{rpj+95}
Rappaport, S., Podsiadlowski, P., Joss, P.~C., DiStefano, R., and Han, Z. 1995,
  MNRAS, 273, 731.

\bibitem[Refsdal and Weigert 1971]{rw71}
Refsdal, S. and Weigert, A. 1971, AA, 13, 367.

\bibitem[Reid 1996]{reid96}
Reid, I.~N. 1996, AJ, 111, 2000.

\bibitem[Rickett 1988]{ric88}
Rickett, B.~J. 1988, in {\em Radio Wave Scattering in the Interstellar Medium
  ,{AIP} Conference Proceedings {N}o. 174}, ed.\ J.M. Cordes, B.~J. Rickett,
  and D.~C. Backer, (New York: American Institute of Physics), 2.

\bibitem[Ruderman 1991]{rud91}
Ruderman, M. 1991, ApJ, 366, 261.

\bibitem[Ryba 1991]{ryb91}
Ryba, M.~F. 1991, PhD thesis, Princeton University.

\bibitem[Ryba and Taylor 1991]{rt91a}
Ryba, M.~F. and Taylor, J.~H. 1991, ApJ, 371, 739.

\bibitem[Sandhu \etal  1997]{sbm+97}
Sandhu, J.~S., Bailes, M., Manchester, R.~N., Navarro, J., Kulkarni, S.~R., and
  Anderson, S.~B. 1997, ApJ, 478, L95.

\bibitem[Sayer 1996]{say96}
Sayer, R.~W. 1996, PhD thesis, Princeton University.

\bibitem[{Schaller} \etal  1992]{ssmm92}
{Schaller}, G., Schaerer, D., Meynet, G., and Maeder, A. 1992, A\&AS, 96, 269.

\bibitem[Schmidt 1968]{sch68}
Schmidt, M. 1968, ApJ, 151, 393.

\bibitem[Shrauner \etal  1996]{ssd+96}
Shrauner, J.~A., Stairs, I.~H., Dewey, R.~J., Krumholz, M., Taylor, H.~E.,
  Taylor, J.~H., and Thorsett, S.~E. 1996, in Johnston \etal 1996, p.\ 23.

\bibitem[Stairs \etal  1998]{sac+98}
Stairs, I.~H., Arzoumanian, Z., Camilo, F., Lyne, A.~G., Nice, D.~J., Taylor,
  J.~H., Thorsett, S.~E., and Wolszczan, A. 1998, ApJ, submitted; 
  astro-ph/9712296.

\bibitem[Stinebring \etal  1992]{skn+92}
Stinebring, D.~R., Kaspi, V.~M., Nice, D.~J., Ryba, M.~F., Taylor, J.~H.,
  Thorsett, S.~E., and Hankins, T.~H. 1992, Rev. Sci. Inst., 63, 3551.

\bibitem[Taam and van den Heuvel 1986]{tv86}
Taam, R. E. and van den Heuvel, E. P. J. 1986, ApJ, 305, 235.

\bibitem[Taylor 1987]{tay87a}
Taylor, J.~H. 1987, in {\em General Relativity and Gravitation}, ed.\ M.~A.~H.
  MacCallum, (Cambridge: Cambridge University Press), 209.

\bibitem[Taylor 1992]{tay92}
Taylor, J.~H. 1992, Phil.\ Trans.\ Roy.\ Soc.\ London A, 341, 117.

\bibitem[Taylor \etal  1995]{tmlc95}
Taylor, J.~H., Manchester, R.~N., Lyne, A.~G., and Camilo, F. 1995, Unpublished
  (available at ftp://pulsar.princeton.edu/pub/catalog).

\bibitem[Taylor and Weisberg 1989]{tw89}
Taylor, J.~H. and Weisberg, J.~M. 1989, ApJ, 345, 434.

\bibitem[Thorsett \etal  1993]{tamt93}
Thorsett, S.~E., Arzoumanian, Z., McKinnon, M.~M., and Taylor,
J.~H. 1993, ApJ, 405, L29.

\bibitem[Thorsett, Arzoumanian, and Taylor 1993]{tat93}
Thorsett, S.~E., Arzoumanian, Z., and Taylor, J.~H. 1993, ApJ,
  412, L33.

\bibitem[Thorsett and Stinebring 1990]{ts90}
Thorsett, S.~E. and Stinebring, D.~R. 1990, ApJ, 361, 644.

\bibitem[van den Heuvel 1994]{vdh94}
van den Heuvel, E. P. J. 1994, A\&A, 291, L39.

\bibitem[van den Heuvel and Bitzaraki 1995]{vb95a}
van den Heuvel, E. P. J., and Bitzaraki, O. 1995, A\&A, 297, L41.

\bibitem[van Kerkwijk 1996]{van96}
van Kerkwijk, M.~H. 1996, in Johnston \etal 1996, p.\ 489.

\bibitem[van Kerkwijk, Bergeron, and Kulkarni 1996]{vbk96}
van Kerkwijk, M.~H., Bergeron, P., and Kulkarni, S.~R. 1996, ApJ, 467, L89.

\bibitem[Webbink, Rappaport, and Savonije 1983]{wrs83}
Webbink, R.~F., Rappaport, S., and Savonije, G.~J. 1983, ApJ, 270, 678.

\bibitem[Wheeler 1966]{whe66}
Wheeler, J.~A. 1966, ARA\&A, 4, 393.

\end{thebibliography}
\end{document}